\documentclass[journal]{IEEEtran}

\IEEEoverridecommandlockouts                              

\usepackage{color}
\usepackage{multirow}
\usepackage{cite}
\usepackage{balance}
\usepackage{enumitem}

\usepackage{color}
\usepackage{multirow}
\usepackage{cite}
\usepackage{balance}
\usepackage{amssymb}
\usepackage{soul}
\usepackage{pifont}
\usepackage{url}

\usepackage[pdftex]{graphicx}
\usepackage[table,xcdraw]{xcolor}
\DeclareGraphicsExtensions{.pdf,.jpeg,.png,.JPG,}
\usepackage{graphicx}

\DeclareGraphicsExtensions{.eps}
\DeclareGraphicsExtensions{.tif}

\usepackage{epstopdf}
\AppendGraphicsExtensions{.tif}

\graphicspath{{figs/}}
\usepackage{epstopdf}
\usepackage{booktabs}

\usepackage{amsmath}
\usepackage{cleveref}
\ifCLASSOPTIONcompsoc
    \usepackage[caption=false, font=normalsize, labelfont=sf, textfont=sf]{subfig}
\else
\usepackage[caption=false, font=footnotesize]{subfig}
\fi

\usepackage{xcolor,cite,etoolbox}
\makeatletter 
\pretocmd\@bibitem{\color{black}\csname keycolor#1\endcsname}{}{\fail}
\newcommand\citecolor[1]{\@namedef{keycolor#1}{\color{blue}}}
\makeatother

\begin{document}

\title{Enhancing Resilience of Distribution Networks by Coordinating Microgrids and Demand Response Programs in Service Restoration}

\author{Ali~Shakeri~Kahnamouei,~\IEEEmembership{Student Member,~IEEE,}
        and Saeed~Lotfifard,~\IEEEmembership{Senior Member,~IEEE}

	\thanks{
		A. Shakeri Kahnamouei, and S. Lotfifard are with the School of Electrical Engineering and Computer Science, Washington State University, Pullman, WA, 99164, USA, (e-mail: a.shakerikahnamouei@wsu.edu, and s.lotfifard@wsu.edu).
	}
}
\maketitle

\begin{abstract}
In case of high impact low probability events, in order to restore the critical loads of the distribution network as much as possible, it is necessary to employ all available resources such as microgrids and distributed generations. This paper presents a two-stage method for critical load restoration after an extreme event utilizing the coordination of all available distributed generations, microgrids, and demand response programs. In the first stage, the post-disaster reconfiguration of the network is determined, and in the second stage, the demand response outputs of the responsive loads and the restoration status of all critical loads are ascertained. In the first stage, an algorithm is proposed to determine electrical islands, and in the second stage, a new model based on the generalized Benders decomposition algorithm is proposed, which considers demand response programs and all operational constraints and aims to maximize the restored critical loads. The effectiveness of the proposed method is verified by simulating the 33-bus and 69-bus test systems.
\end{abstract}


\section*{Nomenclature}
\addcontentsline{toc}{section}{Nomenclature}
\noindent\textit{Indices and Sets}
\begin{IEEEdescription}[\IEEEusemathlabelsep\IEEEsetlabelwidth{$V_1,V_2,V_3$}]
\item[\small{$i,j,p,q$}] Indices of nodes in the distribution system
\item[\small{$d$}] Index of DR blocks
\item[\small{$N$}] Set of all nodes in the post-disaster distribution system
\item[\small{$M$}] Set of nodes with distributed generations
\item[\small{$L$}] Set of nodes with loads
\item[\small{$EMG$}] Set of load nodes in MGs with extra power
\item[\small{$\Lambda$}] Set of nodes of the loads with DR programs
\item[\small{$E$}] Set of all available lines in the post-disaster distribution system
\item[\small{$S_{ij}$}] Set of nodes in the shortest path between load $i$ and $DG_j$
\item[\small{$\Psi(i)$}] Set of neighboring nodes of the node $i$
\item[\small{$\Phi(i)/\Omega(i)$}] Set of parent/child nodes of the node $i$
\item[\small{$D_i$}] Set of DR blocks of the load $i$
\end{IEEEdescription}
\noindent\textit{Parameters}
\begin{IEEEdescription}[\IEEEusemathlabelsep\IEEEsetlabelwidth{$V_1,V_2,V_3$}]
\item[\small{$r_{ij}/x_{ij}$}] Resistance/reactance of the line $(i,j)$
\item[\small{$P_i^{LD}/Q_i^{LD}$}] Active/reactive load at node $i$
\item[\small{$\varphi_i$}] Power factor of the load at node $i$
\item[\small{$Pr_i^L$}] Priority factor of the load at node $i$
\item[\small{$Pr_{id}^{DR}$}] Priority factor of DR block $d$ of the load $i$
\item[\small{$P_i^{LD,pu}$}] Active load at node $i$ in per unit
\item[\small{$P_i^{DG,Max,pu}$}] Maximum capacity of $DG_i$ in per unit
\item[\small{$|Z_{pq}|$}] Electrical distance of the line $(p,q)$ in the shortest path between load $i$ and $DG_j$
\vspace{+0.1cm}
\item[\smash{\begin{IEEEeqnarraybox*}[][t]{l}
\small{V^{Max}}\\
\small{/V^{Min}}
\end{IEEEeqnarraybox*}}] Upper/lower limit of voltage magnitudes\\
\mbox{}
\vspace{+0.05cm}
\item[\smash{\begin{IEEEeqnarraybox*}[][t]{l}
\small{Q_i^{DG,Max}}\\
\small{/Q_i^{DG,Min}}
\end{IEEEeqnarraybox*}}] Upper/lower limit of reactive power of $DG_i$\\
\mbox{}
\vspace{+0.1cm}
\item[\small{$S_i^{DG,Max}$}] Upper limit of apparent power of the $DG_i$
\vspace{+0.05cm}
\item[\small{$I_{ij}^{Max}$}] Maximum current of the line $(i,j)$
\vspace{+0.05cm}
\item[\small{$P_{id}^{DR,b}$}] Capacity of DR block $d$ of the load $i$
\item[\small{$r$}] The root node in spanning tree
\end{IEEEdescription}
\noindent\textit{Variables}
\begin{IEEEdescription}[\IEEEusemathlabelsep\IEEEsetlabelwidth{$V_1,V_2,V_3$}]
\item[\small{$\omega_i^L$}] Binary variable representing status of the load $i$ after CLR
\item[\small{$\omega_{id}^{DR}$}] Binary variable representing status of DR block $d$ of the load $i$
\item[\small{$\alpha_{ij}$}] Binary variable representing status of the line $(i,j)$
\item[\small{$\beta_{ij}$}] Binary variable indicating the parents of nodes, if node $j$ is a parent of node $i$, $\beta_{ij}=1$
\item[\small{$V_i$}] Voltage magnitude at node $i$
\item[\small{$P_i^L/Q_i^L$}] Restored active/reactive load at node $i$
\item[\small{$P_i^G/Q_i^G$}] Generated active/reactive power at node $i$
\item[\small{$P_i^{Inj}/Q_i^{Inj}$}] Injected active/reactive power at node $i$
\item[\small{$P_{ij}/Q_{ij}$}] Active/reactive power flow of the line $(i,j)$
\item[\small{$I_{ij}$}] Flowing current of the line $(i,j)$
\item[\small{$P_i^{DR}$}] Power reduced by DR program at node $i$
\item[\small{$\delta_{id}^{DR}$}] The used portion of DR block $d$ of the load $i$ in CLR
\end{IEEEdescription}

\vspace{-0.1cm}
\section{Introduction}
\label{section_1}
\IEEEPARstart{H}{igh} impact low probability events (HILP) such as natural catastrophes can cause severe destruction to distribution networks, which yields to critical components’ outages and major losses. Hence, enhancing the resilience of distribution networks against such catastrophic events is unavoidable \cite{7489002}. According to \cite{8361387,8974395}, the sequence of resilience construct is robustness, resourcefulness, and rapid recovery.

Critical load restoration (CLR) strategy has been studied in several papers aiming to improve the distribution network's resilience after an extreme event. In \cite{7127029,7017458,7741533,7782459,hussain2019microgrids,8440111,sedgh2021resilient,momen2020using}, the formation of microgrids (MGs) after a disaster has been addressed. In \cite{7127029}, available distributed generations (DGs) and automatic remotely controlled switches are utilized to form MGs after a disaster. Controllable switches can be utilized to determine the optimal reconfiguration of the post-disaster condition. Furthermore, reconfiguration of the distribution network by employing self-supplied MGs has been discussed in \cite{7017458}. Networked MGs and their capabilities are deployed to model the load restoration problem in a distribution network with the objective of restoring the loads as much as possible while minimizing the restoration time \cite{7741533}. Dynamic formation of MGs in post-disaster timeframe with incorporating advanced technologies of MGs, i.e., smart switches, distributed energy resource management system, advanced distribution management system, and microgrid energy management system has been focused in \cite{7782459}, and dynamic formation of self-sustainable and networked MGs in improving power system resilience has been analyzed \cite{hussain2019microgrids}. Also, microgrid formation and load switching sequence steps are comprised in \cite{8440111} to prevent subsequent outages in the extended event after service restoration. In \cite{sedgh2021resilient}, a robust receding horizon recovery strategy is proposed to form multiple MGs to increase the number of restored critical loads. The proposed method updates the system information dynamically. The stored energy in electric vehicles is used to restore critical loads in \cite{momen2020using}. The aggregation of electric vehicles is seen as a power source and is used in microgrid formation. The utilization of DGs to enhance the resilience of distribution networks has been studied in \cite{8016636,ding2017resilient}. A novel method for critical load restoration of secondary network distribution system utilizing DGs has been proposed in \cite{8016636}. It has been assumed that there is a hierarchical control infrastructure with a centralized control scheme and both DGs and loads are controllable, which might not be practical in the present distribution networks. Multiple DGs coordination has been addressed in \cite{ding2017resilient} with a master-slave control technique. Additionally, quantifying the resilience of distribution networks has been discussed, and several indices introduced in \cite{8421054}, and uncertain devices and asynchronous information are considered in the proposed critical load restoration approach in \cite{8771107}.

The role of microgrids in critical load restoration strategies to enhance the resilience of the network is discussed in \cite{8274054,ghasemi2019decision,7513408,lu2016using,arif2017networked,8606920}. The capability of microgrids in preventing power disruptions in presence of extreme events is discussed in \cite{8274054}. In \cite{ghasemi2019decision}, a decision-making critical load restoration strategy by using microgrids is proposed in which the possible paths for post-event reconfiguration of the network are found by Dijkstra's algorithm. The objective function of the proposed method is maximizing the number of restored critical loads while considering the number of switching operations and the unavailability of the restoration paths. However, the ownerships of MGs and the impact of responsive loads are not considered. In a critical load restoration strategy by using microgrids, the stability of microgrids, limits on frequency deviation, and voltage and current of DG are taken into account in \cite{7513408}. Reference \cite{lu2016using} discusses a reliable control architecture to control different devices while using microgrids to enhance the resilience of the network. Also, advanced demand response management methods are proven to be required to deal with the uncertainty of DGs. In \cite{arif2017networked}, a stochastic mixed-integer linear program is proposed as a critical load restoration strategy in which both dispatchable and non-dispatchable DGs are considered as well as the uncertainty of the DGs and loads. The impact of uncertainties on the resilient operation of microgrids is also studied. A data-driven approach is proposed in \cite{8606920} to estimate the uncertainties dynamically based on historical data. Additionally, a resilience-oriented demand response program and a resilience index are proposed to reduce load sheddings and to quantify the resilience of the microgrids, respectively.

Considering demand response programs in microgrids operation is discussed in \cite{7386706,6509996,tabar2019energy}. A responsive load control considering customer comfort and load priorities is proposed in \cite{7386706}. In order to enhance the resilience of microgrids after islanding, innovative functionalities are presented in \cite{6509996} to manage microgrid storage considering the incorporation of demand response programs. Electrical energy storage systems have been used as the transient power supplier in \cite{tabar2019energy} to tackle the problems caused by shiftable loads. Also, the functionality of demand response programs, as well as demand costs and payments, are considered in the operation of microgrids.

Furthermore, research on new algorithms to solve the problem to increase the resilience of distribution networks and decrease the computational burden of the problem has been conducted in \cite{8616854}. Graph theory and spanning tree research algorithm has been used to create a novel restoration strategy.

Microgrid formations after an extreme event and coordinating local sources within a microgrid have been studied in the literature. Coordinating the resources in all microgrids within an electrical island is not fully addressed \cite{8606281} and utilizing demand response (DR) programs to increase the possible restored critical loads and respecting the ownership of the microgrids within an electrical island are not sufficiently addressed yet. Fig. \ref{fig_1} illustrates the coordination of MGs and responsive loads with respect to microgrids’ ownerships in critical load restoration.

In this paper, a two-stage method is proposed, which determines the reconfiguration of the distribution network at the first stage, and the restored critical loads, the optimal output of DGs in MGs, and the output of the responsive loads are determined at the second stage. The second stage constitutes a mixed-integer nonlinear problem (MINLP), which solving this stage utilizing mathematical programming is challenging. Additionally, heuristic and metaheuristic have been used to solve MINLP problems. In this study, generalized Benders decomposition (GBD) is used to solve the convexified model of the proposed MINLP model in the second stage. GBD is a two-stage optimization strategy where in the first stage, an MILP is solved to determine the status of the loads and demand responses, and in the second stage, the optimal power flow is solved. The novel contributions of the paper are listed as follows:

\begin{enumerate}
    \item The proposed critical load restoration strategy takes into account the ownership of the available microgrids in the distribution system. Unlike previous CLR methods, which do not consider the microgrids' ownership, in the proposed CLR method, a microgrid is considered to have a private owner or a DSO property. Since in a more realistic distribution network, a private MG might not be willing to participate in the critical load restoration, and its priority is supplying its own critical loads rather than helping to restore the critical loads of the other MGs or distribution network, the ownership of the MGs is considered in the proposed CLR method.
    \item A linearized incentive-based demand response model is proposed to enable the participation of loads of the private microgrids with extra power generation in the coordination of the MGs in the CLR. The low priority loads of the private MGs with extra generation can have DR contracts so the DSO can curtail their loads and restore more critical loads located in the other MGs or the rest of the distribution network. In other words, the DSO cannot curtail the low priority loads of a private microgrid without having a demand response contract. Moreover, since the loads offering DR contracts are in the private MGs, in addition to the low priority loads, medium priority loads might offer DR contracts as well to participate in the CLR strategy. However, in the test cases, it is assumed that medium priority loads are not participating in the DR process.
    \item A flowchart is proposed to detect the electrical islands after an HILP event. Detecting the possible electrical islands can enhance the allocation of the survived DGs, resulting in restoring more critical loads. Additionally, the benefits of coordinating microgrids within electrical islands are listed in section \ref{section_2_2}.
    \item After detecting the electrical islands, a two-stage method for CLR is proposed, which considers all available DGs, private and non-private MGs, and the introduced incentive-based DR programs. At the first stage, a novel heuristic is proposed to determine the post-event topology of the distribution network. Also, a GBD-based approach is utilized to solve the proposed mixed-integer second-order cone-programming (MISOCP) problem of the second stage, which alleviates the computational burden of the problem.
\end{enumerate}

The remainder of this paper is organized as follows. Section \ref{section_2} describes the framework of the proposed method, including microgrids coordination and demand response programs. In section \ref{section_3}, the flowchart of determining electrical islands and a heuristic to determine the post-disaster topology of the electrical islands is presented. Furthermore, the problem formulation of the second stage is represented. In section \ref{section_4}, the two-stage GBD method is presented, and in section \ref{section_5}, the results of the case study are presented. Finally, the conclusion and future work is presented in section \ref{section_6}.

\vspace{-0.1cm}
\section{Coordinating Microgrids and Demand Response Programs}
\label{section_2}
This section describes the proposed framework for determining the electrical islands, coordinating local resources of microgrids within an electrical island considering the ownership of the microgrids, and the effectiveness of demand response programs on service restoration.

\vspace{-0.2cm}
\subsection{Assumptions}
\label{section_2_1}
There are several assumptions in the proposed critical load restoration method which should be considered. The assumptions are as follows.

\begin{enumerate}
    \item It is assumed that after a drastic incident, the substations are in outage mode, and the distribution network cannot be supplied by the main grid anymore. Due to the loss of transmission lines and some critical facilities of the distribution network, the distribution system could rely just on local resources and existing microgrids.
    \item After the event, the protection system of the distribution network isolates the faulted areas immediately. Microgrids open their switches and supply their own loads with their survived local DGs.
    \item Generally, microgrids can have private owners, or they can be a property of a distribution system operator (DSO). In other words, unlike most of the previous studies, we do not ignore the ownership of microgrids. Nevertheless, determining the details of contracts that MGs should have with DSO to perform in the post-disaster condition is not in the scope of this paper.
    \item The priority of privately-owned microgrids is to supply their own loads in the post-disaster timeframe. Therefore, each private MG with extra generation which participates in the restoration program which is supposed to curtail some of its low priority loads and provide its generation for DSO to use in service restoration to restore more critical loads, should have DR contracts with its low priority loads and curtail them according to their contracts.
    \item The distribution system operator aims to restore the critical loads of the distribution network as much as possible regardless of the loads' location. If a microgrid does not participate in the restoration program, its switch would be open, and the DSO would not consider the presence of critical loads in that specific MG.
\end{enumerate}

\begin{figure}[t]
\centering
\includegraphics[width=0.95\linewidth]{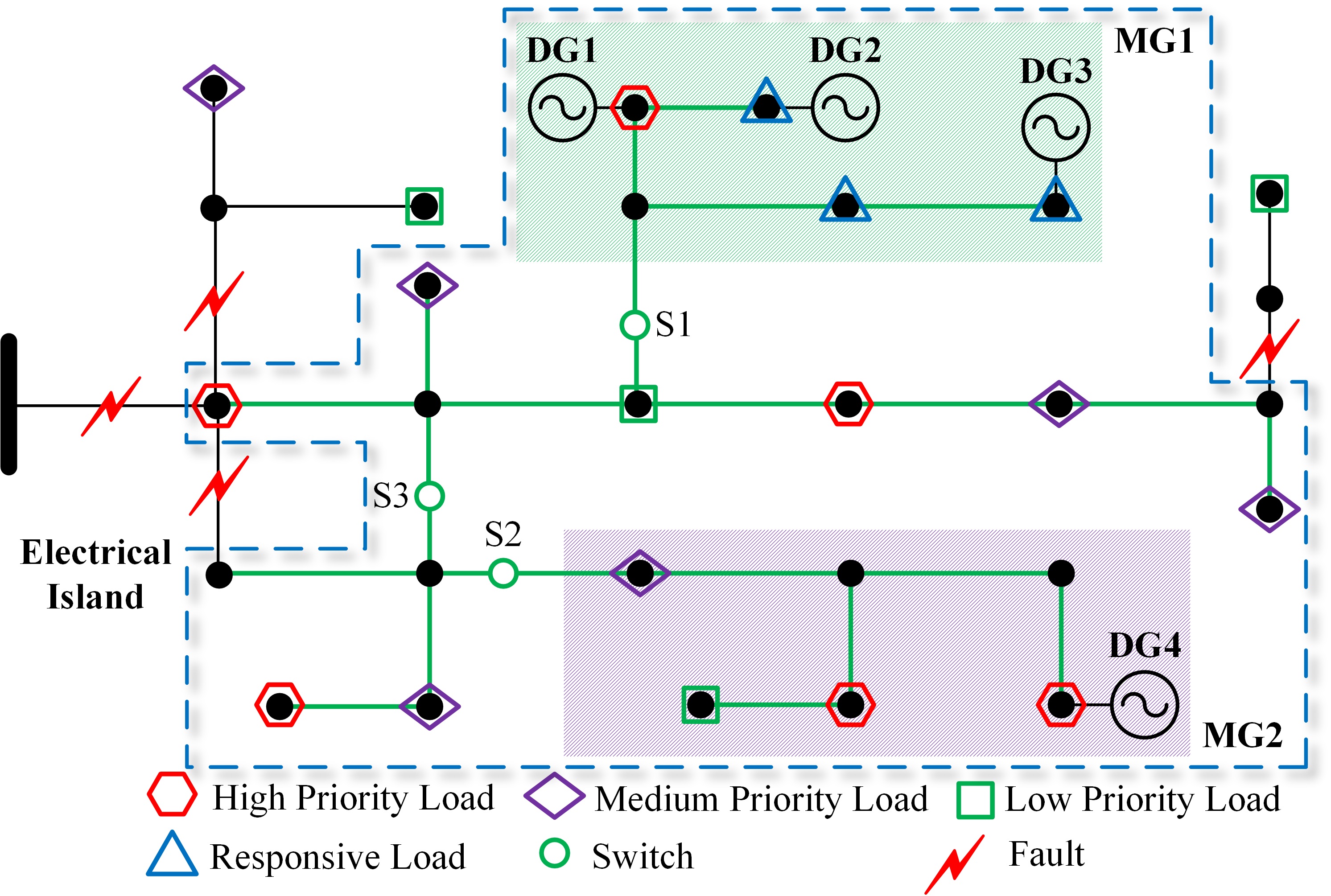}\vspace{-0.2cm}
\caption{Coordinating MGs considering DR programs in CLR}
\vspace{-0.5cm}
\label{fig_1}
\end{figure}

\vspace{-0.2cm}
\subsection{Microgrids Coordination}
\label{section_2_2}
After an extreme event, due to the opening of some switches in the distribution system, several electrical islands can be formed \cite{6999983}. By further interconnecting these islands by closing possible survived tie lines and switches, larger islands with more local resources can be formed. Consequently, multiple sources, including any kinds of DGs in different MGs, can be coordinated to determine the optimal allocation of available resources for critical load restoration. In Fig. \ref{fig_1}, in order to coordinate the MGs, their switches, i.e., $s_1$, and $s_2$, are closed. Then, by closing $s_3$, a larger electrical island can be formed. Microgrid with extra power (MG1) has DR contracts with its low priority loads. The main benefit of these DR programs is that instead of serving low priority loads of MG1, DSO can use the extra power of MG1 to supply critical loads in the distribution network or MG2. Coordinating microgrids within the electrical islands can have some benefits \cite{8606281}.

\begin{enumerate}
    \item The possibility of supplying more critical loads will be increased by pooling resources with low capacities.
    \item Optimal allocation of DGs with limited generation can yield to increase the duration of supplying critical loads. In other words, the total time in which critical loads are supplied is influenced by the location of the DGs in the microgrids.
    \item By coordinating the control features of multiple resources, the robustness of the distribution network can be enhanced, resulting to better withstand extreme events during service restoration.
    \item Managing uncertainties of renewable energy resources with diverse operational characteristics can be improved by their coordinated performance.
\end{enumerate}

\vspace{-0.2cm}
\subsection{Demand Response Programs}
\label{section_2_3}
As the priority of the loads in a distribution system is different, the price of power for every load is different. Consequently, DSO might be faced with significant penalties due to curtailing critical loads in the distribution systems. The utilization of demand response programs is one of the economical, reliable, and resilient methods to not only avoid these penalties but also enhance the resiliency of the distribution network.

Conventionally, in the presence of reliability problems in a distribution system or high prices of the electricity market, end-use customers take some short-term actions in order to reduce their electricity use, which is called demand response \cite{FERK}. Demand response impacts on the resilience of a distribution system have been analyzed in several studies \cite{mousavizadeh2018linear}. According to \cite{FERK}, there are two major categories of DR programs: i) Incentive-based programs (IBP), and ii) price-based programs (PBP). An objective-wised classification of DR programs has been proposed in \cite{6039608}, and it has been concluded that the majority of DR contracts in the US markets are used to enhance the reliability and resiliency of the distribution networks. Since our goal is utilizing DR programs in order to increase the resiliency of the distribution system, the types of DR contracts should be incentive-based. The incentive-based DR programs’ sub-categories are as follows \cite{5874702}:
\subsubsection{Direct load control (DLC)}
\label{section_2_3_1}
In DLC programs, incentive payments are offered to the customers to allow the DSO to curtail their loads on short notice in the presence of the reliability contingencies. Typically, these programs are operated during the system peak demand.
\subsubsection{Interruptible/curtailable service (I/C)}
\label{section_2_3_2}
In I/C services, customers are responsible for reducing their loads in the presence of the contingencies in exchange for a bill credit or rate discount. In I/C-based contracts, some penalties are considered for the customers in case of not reducing their loads.
\subsubsection{Emergency demand response program (EDRP)}
\label{section_2_3_3}
In EDRP programs, customers are paid to reduce their loads during reliability-related or resiliency-oriented incidents. These incentive payments are determined in the contracts prior to extreme events.
\subsubsection{Capacity market program (CMP)}
\label{section_2_3_4}
The CMP programs are similar to insurance programs. Customers receive guaranteed payments and are agreed to reduce a pre-determined amount of their loads while contingencies occur. The customers might not be asked to curtail their loads if no contingency happens. Some penalties are also considered for the customers if they decline to reduce their loads during the contingencies.
\subsubsection{Demand bidding/buyback}
\label{section_2_3_5}
In this DR program, large customers offer to curtail their loads at a pre-specified price. Therefore, if customer bids are cheaper than the market, they will be asked to curtail their loads. Large customers usually are interested in this program.
\subsubsection{Ancillary service markets (A/S)}
\label{section_2_3_6}
A/S program refers to the operation reserves that are conducted by customers' offers to curtail their loads. If the customers’ bids are accepted, in case they are called to reduce their loads, they will be paid the spot market price. The ability to rapid adjusting of the load is a necessity for customers to be able to participate in this program. Unlike peak shaving DR programs, which need the hourly response of the customers to curtail their loads, the response time in this program is in the range of minutes.

In \cite{5677457}, a pool-based demand response exchange (DRX) model is proposed in which all operations of buyers and sellers are managed by the demand response exchange operator (DRXO). However, this method does not consider the available microgrids and their ownerships. Additionally, the market-clearing model of the proposed DRX model is not applicable in the proposed decision-making CLR by coordinating microgrids. To address this challenge, an incentive-based DR program is proposed.

\begin{figure}[t]
\centering
\includegraphics[width=0.7\linewidth]{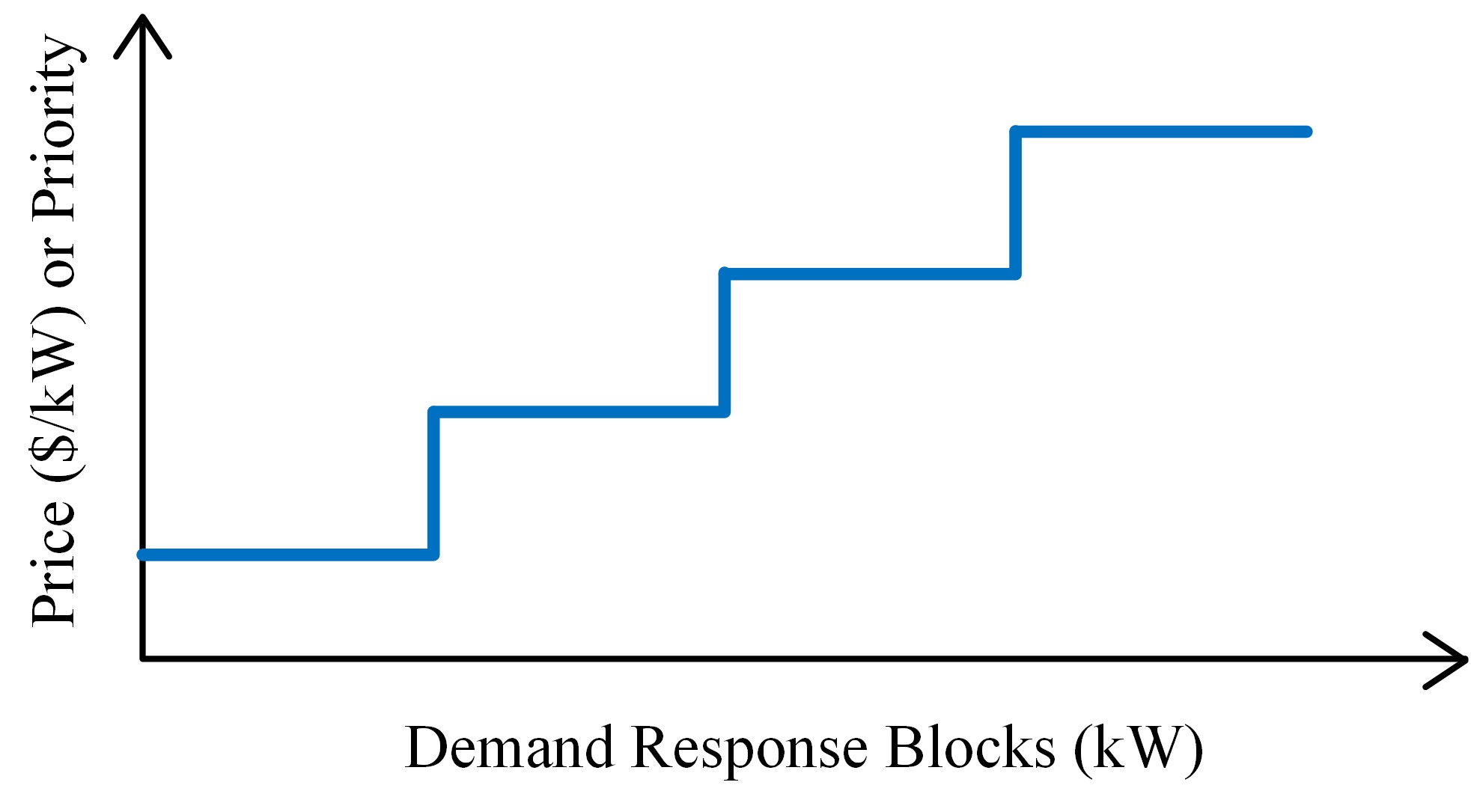}\vspace{-0.2cm}
\caption{Used demand response programs in CLR}
\vspace{-0.5cm}
\label{fig_2}
\end{figure}

In this paper, the low priority loads of the private MGs with extra generation are paid within a multi-step EDRP program contract with DSO to curtail their loads in presence of an HILP event. Each step includes a specific percentage of the total load. The amount of offered DR in all the steps is not necessarily equal. The price of DR increases in each step, meaning that the first step has the lowest price. Conceptually discussing, the price of the last 1 kW curtailed load is more expensive than the first 1 kW. Since the proposed method is decision-making, instead of considering prices of DR blocks, their priorities have been utilized. In other words, the problem is deciding whether a load should be supplied or curtailed. Fig. \ref{fig_2} represents the proposed demand response program, which is utilized in the proposed critical load restoration. The y-axis represents the priorities of the blocks, which is equivalent to their prices.

\vspace{-0.1cm}
\section{Problem Formulation}
\label{section_3}
This section describes the proposed strategy for determining the electrical islands, coordinating local resources of microgrids within an electrical island considering the ownership of the microgrids, and the effectiveness of demand response programs on service restoration.

\vspace{-0.2cm}
\subsection{Detecting Electrical Islands}
\label{section_3_1}
After a disastrous event, the first step is detecting the possible electrical islands. All MGs and loads in the distribution system should be in the same electrical island unless they are entirely isolated, and there are no switches. By aggregating and coordinating all possible MGs in an electrical island, the allocation of the available DGs in MGs can be enhanced, and therefore more critical loads can be restored. The flowchart of the proposed method to detect electrical islands is represented in Fig. \ref{fig_3}. Set $M$ presents all survived DGs in the distribution network. $M^*$ is an auxiliary set, which takes all DGs of $M$ and puts each DG in an electrical island, which should be empty at the end. $i$ and $j$ are indices for DGs, and $k$ is the index of electrical islands. If there is any possible path between nodes containing $DG_i$ and $DG_j$, including lines with open and closed switches, then they should be in the same electrical island. The same algorithm is utilized to assign the load nodes to electrical islands.

\vspace{-0.2cm}
\subsection{Determining the Post-Disaster Topology in the First Stage}
\label{section_3_2}
In the first stage, a new method is presented to determine the topology of the electrical islands. After detecting the target islands as electrical islands by the method introduced in the previous section, the topology of each electrical island will be determined.

In order to determine the radial post-disaster topology of electrical islands, a method based on graph theory is proposed in \cite{8606281}. In this paper, a new heuristic to determine the post-disaster topology of electrical islands is proposed, which combines graph theory with the specific characteristics of the distribution network after an extreme event. The motivation for proposing this method is considering the resilience of the distribution network, even in detecting the post-disaster topology. In other words, the proposed method considers the possible aftershocks and increases the probability of supplying critical loads. It should be noted that the proposed method works for any kinds of faults and outages caused by HILP events. The objective of the CLR strategy is to serve the critical loads during the time that the main power grid is unavailable. The power of the main grid will be back after several hours or days, depending on the severity of the event and damaged components. Furthermore, after natural disasters such as earthquakes, the distribution system, and post-disaster topology is vulnerable to aftershocks. Thus, the resiliency of the distribution system could be considered in determining the post-event topology to increase the probability of serving the restored critical loads during the post-event timeframe. Since the main power grid is unavailable, it is assumed that the total load of the distribution system is more than the available capacity of DGs. Hence, some loads, including some critical loads, cannot be served. Consequently, the post-event topology should provide the path, i.e., lines such that critical loads with high power consumption can have the least possible electrical distance from DGs with high capacity. In other words, the less electrical distance between critical load and DGs, the less probability of losing the path. Then, the idea is that the electrical distance of the critical loads from DGs should be minimal. In order to find the shortest path between each load and DG, the so-called $A^*$ algorithm is used. The goal is to maximize the weighted binary status of the lines in the paths considering the radiality constraints of the distribution network. The loads and capacity of the DGs can be normalized. Instead, the per-unit values are used. Therefore, the objective function of the first stage is defined as follows:

\begin{figure}[t]
\centering
\includegraphics[width=0.98\linewidth]{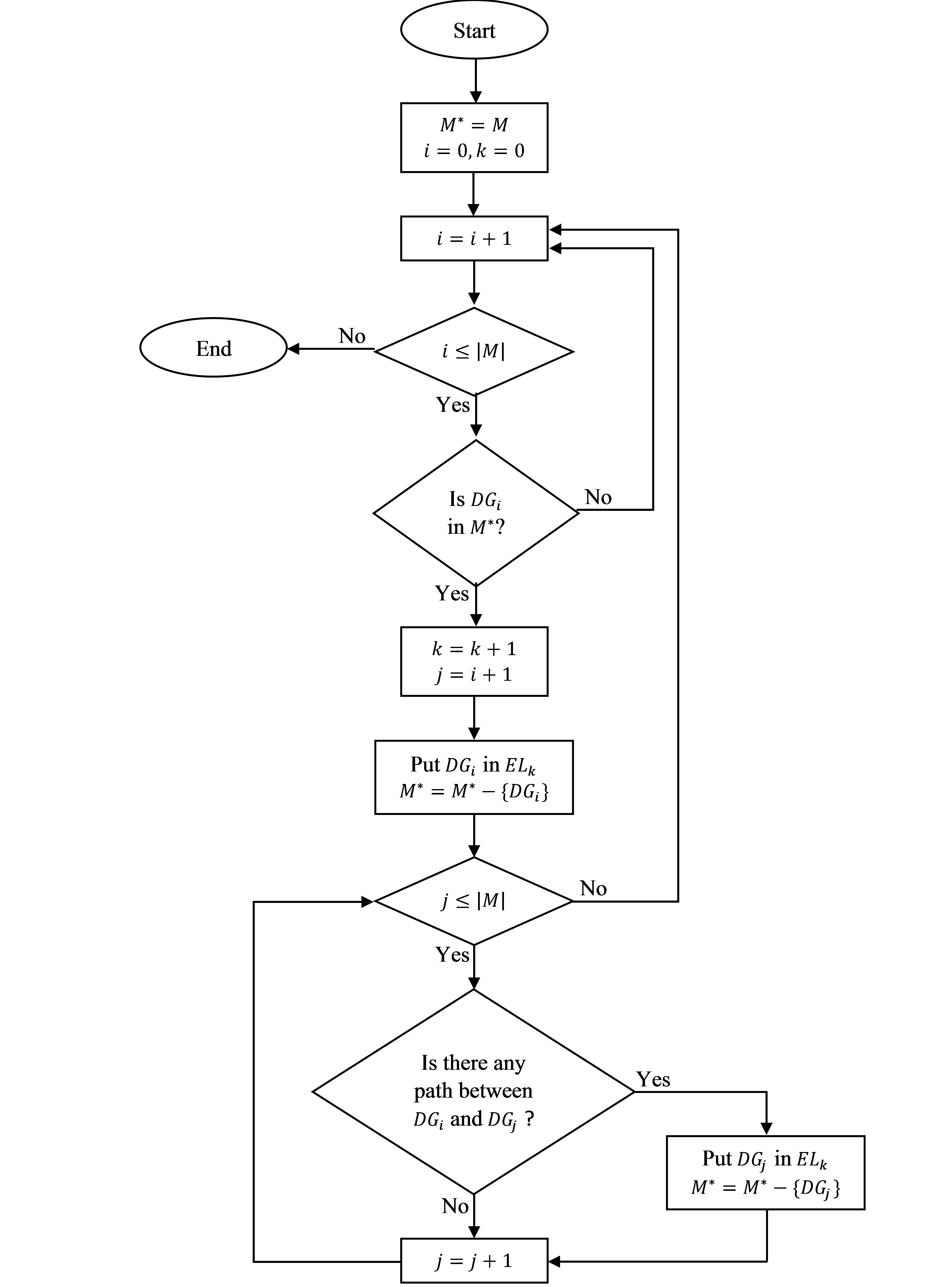}\vspace{-0.2cm}
\caption{The flowchart of detecting electrical islands}
\vspace{-0.5cm}
\label{fig_3}
\end{figure}

\vspace{-0.03cm}
\begin{equation}
\label{eq_1}
\small
\max{ \sum_{i \in L}{ \sum_{j \in M}{ Pr_i^L P_i^{LD,pu} d_{ij} P_j^{DG,Max,pu} } } }
\end{equation}
where $d_{ij}$ is defined as:
\vspace{-0.03cm}
\begin{equation}
\label{eq_2}
\small
d_{ij} = \frac{1}{2} \sum_{p \in S_{ij}}{ \sum_{q \in S_{ij}}{ \alpha_{pq} |Z_{pq}| } }, \quad\forall{ i \in L, j \in M }
\end{equation}
Since each line is counted twice, the equation is divided by two. Radiality, i.e., spanning tree constraints \cite{6153415} for each electrical island, are represented in (\ref{eq_3})-(\ref{eq_5}).
\vspace{-0.03cm}
\begin{equation}
\label{eq_3}
\small
\beta_{ij} + \beta_{ji} = \alpha_{ij}, \quad\forall{ (i,j) \in E }
\end{equation}
\vspace{-0.3cm}
\begin{equation}
\label{eq_4}
\small
\sum_{j \in \Psi(i)}{\beta_{ij}} = 1, \quad\forall{ i \in N \backslash r }
\end{equation}
\vspace{-0.3cm}
\begin{equation}
\label{eq_5}
\small
\beta_{ij} = 0, \quad\forall{ i \in r, j \in \Psi(i) }
\end{equation}

\begin{figure}[t]
\centering
\includegraphics[width=0.6\linewidth]{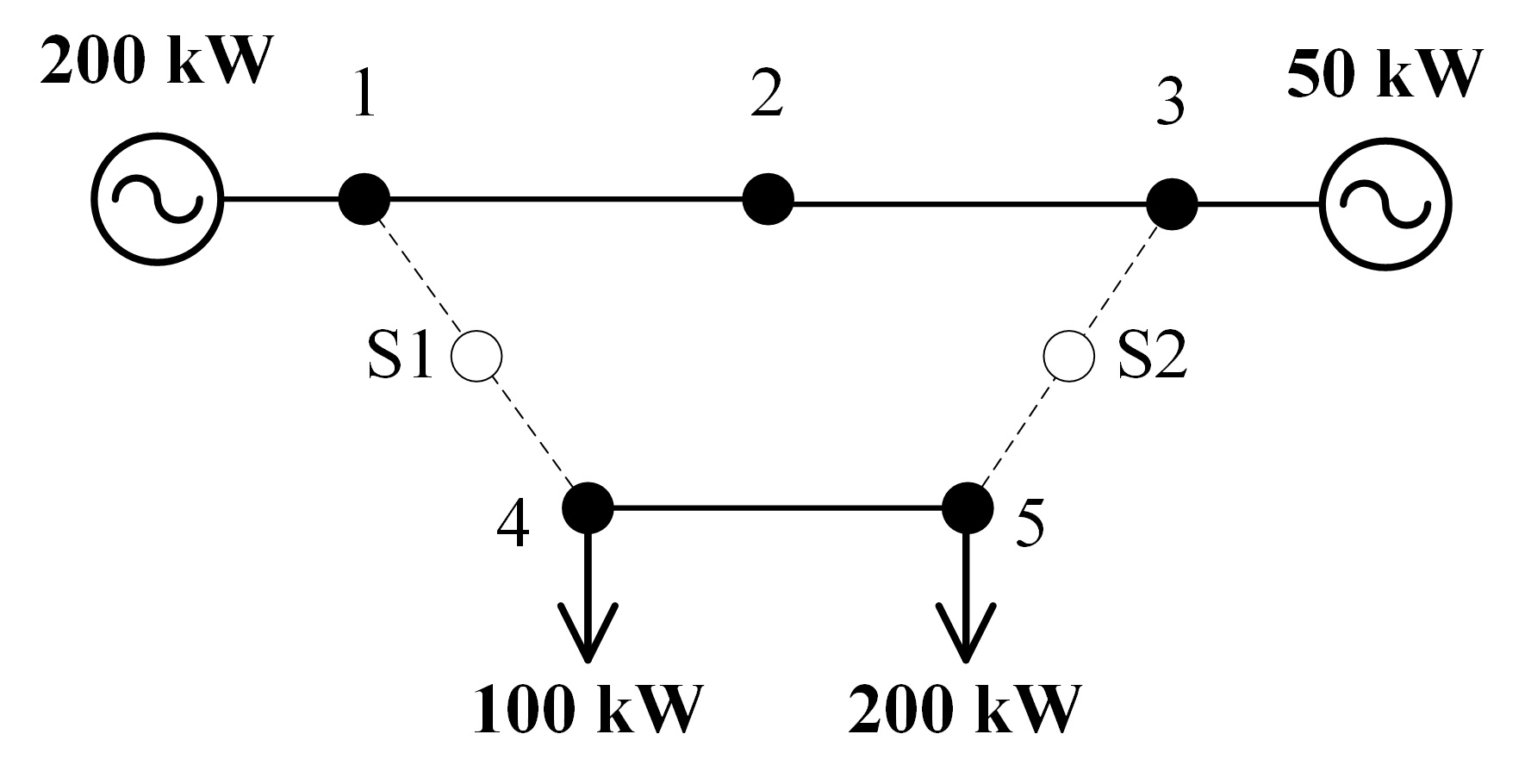}\vspace{-0.2cm}
\caption{An instance to detect the topology of an electrical island}
\vspace{-0.5cm}
\label{fig_4}
\end{figure}

Fig. \ref{fig_4} illustrates the proposed method with an electrical island. Total loads of the electrical island are more than the total capacity of DGs. This is the main assumption to prove the validity of the proposed method in most cases. Indeed, if the total capacity of DGs was more than the total load, the proposed method might not be the preferred method considering line losses and voltage profile. Since the main power grid is not available, the assumption is highly acceptable. Suppose the electrical distance of each line to be 1 unit. Node 1 is the root node. It should be noted that at each electrical island, the DG with higher power generation capacity is selected as the dominant power generator, which is the reference bus to set the reference voltage. Choosing the DG with higher generation capacity has several advantages, such as having a better voltage profile of the electrical island. In order to satisfy the radiality constraints, one of the switches of $s_1$ or $s_2$ should be closed. Both loads in nodes 4 and 5 are supposed to be critical. Thus, their priorities are the same. The load in node 5 is twice greater than the load in node 4. Therefore, while maximizing the objective function, the shortest path between the greater load (node 5) and the DG with high capacity (node 1) will be energized. Consequently, the $s_1$ switch will be closed, and the load at node 5 will be restored. Comparing two possible topologies, if $s_2$ was supposed to be closed instead of $s_1$, since a significant portion of the load in node 5 is supplied by the DG in node 1, the power would be delivered through a longer electrical distance, resulting in more line losses and reduced voltage profile. Additionally, the resiliency of the topology would be reduced as in that case, 4 energized lines are needed to supply the restored critical load, which is twice more than the first topology. Furthermore, since the electrical island is in one geographical area, the probability of lines being affected by an extreme event is equal.

The proposed method in \cite{8606281} is a graph-theoretic algorithm to find the post-restoration topology of the network. In this method, the minimum diameter spanning tree (MDST) is selected as the post-fault topology without considering the loads and survived DGs of the network. However, in the proposed heuristic algorithm to find the post-fault topology of the network, alongside utilizing the graph theory, the loads and survived DGs of the network are also considered by introducing equations (\ref{eq_1})-(\ref{eq_2}). Consequently, the proposed algorithm is more applicable.

\vspace{-0.2cm}
\subsection{CLR Using DR Programs in the Second Stage}
\label{section_3_3}
After determining the topology of the electrical island in the first stage, the critical load restoration strategy should be applied to restore the critical loads as much as possible. In this section, the problem formulation for critical load restoration by utilizing demand response programs is presented. The objective function is to maximize the weighted restored loads while minimizing the responsive loads. The objective function of the second stage is introduced as follows:
\vspace{-0.03cm}
\begin{equation}
\label{eq_6}
\small
\max{ \sum_{i \in L}{ Pr_i^L \omega_i^L } - \sum_{i \in \Lambda}{ \sum_{d \in D_i}{ Pr_i^L Pr_{id}^{DR} \omega_{id}^{DR} } } }
\end{equation}

The preference of the responsive loads is considered using the priority factors ($Pr_i^L$). Also, each responsive load has several DR blocks with different priority factors ($Pr_{id}^{DR}$).

In this paper, the second-order cone-programming (SOCP) model as a convex model of distribution system reconfiguration has been used \cite{6153415,7432265}. The reason to use the SOCP model is that not only the operational time is significantly reduced, but also since it is convex, the global optimum point can be achieved. Power flow constraints of the CLR problem in the second stage are written based on the DistFlow branch model for radial distribution networks as follows \cite{7432265}:
\vspace{-0.03cm}
\begin{equation}
\label{eq_7}
\small
\sum_{j \in \Phi_i}{(P_{ji} - r_{ji}I_{ji}^2)} + P_i^{Inj} = \sum_{k \in \Omega_i}{P_{ik}}, \quad\forall{ i \in { N \backslash r } }
\end{equation}
\vspace{-0.3cm}
\begin{equation}
\label{eq_8}
\small
\sum_{j \in \Phi_i}{(Q_{ji} - x_{ji}I_{ji}^2)} + Q_i^{Inj} = \sum_{k \in \Omega_i}{Q_{ik}}, \quad\forall{ i \in { N \backslash r } }
\end{equation}
\vspace{-0.3cm}
\begin{equation}
\label{eq_9}
\small
P_i^{Inj} = P_i^G - P_i^L, \quad\forall{ i \in N }
\end{equation}
\vspace{-0.3cm}
\begin{equation}
\label{eq_10}
\small
Q_i^{Inj} = Q_i^G - Q_i^L, \quad\forall{ i \in N }
\end{equation}
\vspace{-0.3cm}
\begin{equation}
\label{eq_11}
\small
I_{ij}^2 = \frac{ P_{ij}^2 + Q_{ij}^2 }{V_i^2}, \quad\forall{ (i,j) \in E }
\end{equation}
\vspace{-0.3cm}
\begin{equation}
\label{eq_12}
\small
\begin{split}
V_i^2 - V_j^2 - 2( r_{ij}P_{ij} + x_{ij}Q_{ij} ) + ( r_{ij}^2 + x_{ij}^2 )I_{ij}^2\\
= 0, \quad\forall{ (i,j) \in E }
\end{split}
\end{equation}
\vspace{-0.3cm}
\begin{equation}
\label{eq_13}
\small
(V^{Min})^2 \leq V_i^2 \leq (V^{Max})^2, \quad\forall{ i \in N }
\end{equation}
\vspace{-0.3cm}
\begin{equation}
\label{eq_14}
\small
0 \leq I_{ij}^2 \leq (I_{ij}^{Max})^2, \quad\forall{ (i,j) \in E }
\end{equation}
\vspace{-0.3cm}
\begin{equation}
\label{eq_15}
\small
(P_i^G)^2 + (Q_i^G)^2 \leq (S_i^{DG,Max})^2, \quad\forall{ i \in M }
\end{equation}
\vspace{-0.3cm}
\begin{equation}
\label{eq_16}
\small
Q_i^{DG,Min} \leq Q_i^G \leq Q_i^{DG,Max}, \quad\forall{ i \in M }
\end{equation}
The active and reactive line flow equations are represented by constraints (\ref{eq_7})-(\ref{eq_8}), respectively. Constraints (\ref{eq_9})-(\ref{eq_10}) are the nodal active and reactive power flows, respectively. The current magnitude of each line is determined by constraint (\ref{eq_11}) and constraint (\ref{eq_12}) represents the ohm’s law over each line. The bus voltage magnitude and line currents are limited by constraints (\ref{eq_13})-(\ref{eq_14}). Constraint (\ref{eq_15}) limits the generated apparent power to its maximum value for each bus. The generated reactive power of each bus is also bounded by its minimum and maximum limits in constraint (\ref{eq_16}).

In the above-mentioned formulation, let replace $V_i^2$ and $I_{ij}^2$ with $v_i$ and $i_{ij}$, respectively. Therefore, the nonlinearities caused by the square of voltage and current magnitudes are linearized. The only remaining nonlinear constraints are (\ref{eq_11}) and (\ref{eq_15}). The equality constraint (\ref{eq_11}) can be relaxed into the following inequality constraint \cite{6756976}:
\vspace{-0.03cm}
\begin{equation}
\label{eq_17}
\small
I_{ij}^2 \geq \frac{ P_{ij}^2 + Q_{ij}^2 }{V_i^2}, \quad\forall{ (i,j) \in E }
\end{equation}

An electrical island consists of several MGs with DGs and loads, and the rest of the distribution system, which does not belong to any MG. An electrical island can be a single-microgrid, multi-microgrid, or no-microgrid island. A single-microgrid island includes an MG in which there is no path to connect the MG to the rest of the system. A multi-microgrid island includes multiple microgrids in which they are networked to supply the critical loads of the island. Also, a no-microgrid island contains one or more DGs that are not in any MGs and there is no path to connect them to any MGs. Therefore, they form an electrical island with no MGs. The MGs could belong to private owners, and any MG which is not willing to participate in the critical load restoration will be islanded and will not be in the electrical island. In other words, if an MG does not participate in the CLR program, its switch will be open, and the MG will be totally isolated from the grid. Hence, it cannot be included in any electrical island. Generally, some of the MGs in the electrical island have private owners. The MGs that do not have private owners and their loads do not have any specific priority over the rest of the distribution system, will not participate in DR programs. The loads that do not belong to any privately-owned MGs do not have any DR contracts, and their importance is shown as a priority coefficient in the formulation. The preference of the private MGs in the electrical island is to serve their own loads rather than helping the DSO to restore the rest of the distribution system's critical loads. It should be noted that the privately-owned MGs, which were dependent on the main power grid, require the DSO’s help to restore their loads. Therefore, they will not have DR contracts, and in the view of DSO, their DGs and loads are like the rest of the distribution system. The DSO utilizes their DGs in the CLR program and tries to restore their critical loads. Consequently, the private MGs with more generation capacity than loads will have DR contracts. Obviously, the critical loads of these MGs will not have any DR contracts because they are the first priority of these MGs to be served.

\begin{figure}[t]
\centering
\includegraphics[width=0.98\linewidth]{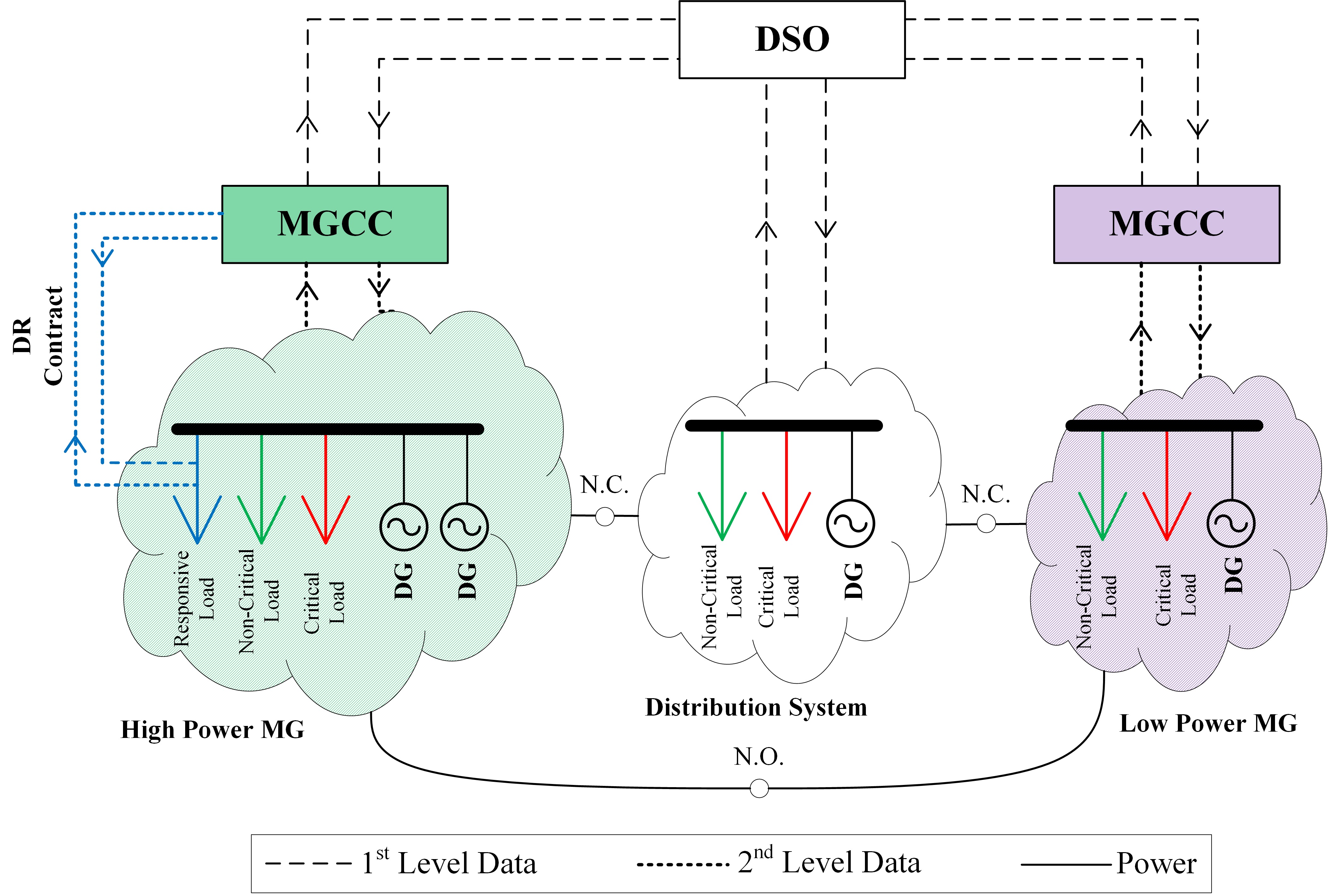}\vspace{-0.2cm}
\caption{The conceptual framework of the proposed method}
\vspace{-0.5cm}
\label{fig_5}
\end{figure}

The conceptual framework of the proposed method in the second stage is represented in Fig. \ref{fig_5}. The distribution network consists of the MGs with extra power (light green block), MGs with low power (light purple block), and the rest of the feeders (white block). The radiality constraints of the system should be satisfied while connecting the MGs to the system. This is shown by normally closed (N.C.) and normally open (N.O.) switches. The microgrids include critical and non-critical loads. Each MG has a microgrid control center (MGCC), which communicates with the DSO. The interactions among the MGs are through the DSO as a centralized control center. Hence, the microgrids control centers do not perform as decentralized control centers, and they need to communicate with the DSO. As represented in Fig. \ref{fig_5}, there is no peer-to-peer platform between the microgrids, and they are not able to communicate independently. Therefore, any data transformation requires to be done through the DSO. The low priority loads of the MG with extra power can have DR contracts. Therefore, their loads can be curtailed in the critical load restoration strategy. Hence, the extra generation of the MG can be used to supply the critical loads of the other MGs and the rest of the distribution system.

Additionally, it should be noted that when an extreme event happens, besides damaging the distribution network, microgrids, DGs, and other facilities of the system, the communication system can also be affected. Hence, the role of the communication system in service restoration is unavoidable. As shown in Fig. \ref{fig_5}, the communication links between the DSO and microgrid control centers and the DGs in the other feeders are essential tools to transfer data. Consequently, if a communication link between the DSO and an MG is damaged, the microgrid cannot be able to participate in the critical load restoration process.

The status of all the loads in the distribution system which do not have DR contracts can be 0 or 1. In other words, they either are restored or not. Hence, there is no partial restoration for these loads. The status of these loads is represented as follows:
\vspace{-0.03cm}
\begin{equation}
\label{eq_18}
\small
P_i^L = \omega_i^L P_i^{LD}, \quad\forall{ i \in { N \backslash \Lambda } }
\end{equation}
\vspace{-0.3cm}
\begin{equation}
\label{eq_19}
\small
Q_i^L = \omega_i^L Q_i^{LD}, \quad\forall{ i \in { N \backslash \Lambda } }
\end{equation}

As stated previously, the critical loads of private MGs with extra generation capacity must be restored. Therefore, the status of these loads should be 1.
\vspace{-0.03cm}
\begin{equation}
\label{eq_20}
\small
\omega_i^L = 1, \quad\forall{ i \in { EMG \backslash \Lambda } }
\end{equation}
The loads having DR contracts may be restored partially. The DR contracts are applied in the following constraints.
\vspace{-0.03cm}
\begin{equation}
\label{eq_21}
\small
P_i^L = P_i^{LD} - P_i^{DR}, \quad\forall{ i \in \Lambda }
\end{equation}
\vspace{-0.3cm}
\begin{equation}
\label{eq_22}
\small
Q_i^L = Q_i^{LD} - tan(\varphi_i) P_i^{DR}, \quad\forall{ i \in \Lambda }
\end{equation}

Equation (\ref{eq_21}) shows the amount of restored active power of the loads with DR contracts. Equation (\ref{eq_22}) indicates that restored reactive power of the loads with DR contract will be reduced in conjunction with the power factor of the load.

Demand response programs are defined in $d$ steps. The capacity of each step is denoted by $P_{id}^{DR,b}$. The total DR contract of each load cannot be greater than its active power consumption. Therefore, the summation of all DR blocks of a load should not be greater than the forecasted active power of the load.
\vspace{-0.03cm}
\begin{equation}
\label{eq_23}
\small
\sum_{d \in D_i}{ P_{id}^{DR,b} } \leq P_i^{LD}, \quad\forall{ i \in \Lambda }
\end{equation}

The amount of applied DR in CLR for each load will be the summation of a portion of each DR block as follows.
\vspace{-0.03cm}
\begin{equation}
\label{eq_24}
\small
P_i^{DR} = \sum_{d \in D_i}{ \omega_{id}^{DR} \delta_{id}^{DR} }, \quad\forall{ i \in \Lambda }
\end{equation}
To linearize constraint (\ref{eq_24}), it can be replaced with the following constraints:
\vspace{-0.03cm}
\begin{equation}
\label{eq_25}
\small
P_i^{DR} = \sum_{d \in D_i}{ \gamma_{id}^{DR} }, \quad\forall{ i \in \Lambda }
\end{equation}
\vspace{-0.3cm}
\begin{equation}
\label{eq_26}
\small
\gamma_{id}^{DR} \leq P_{id}^{DR,b} \omega_{id}^{DR}, \quad\forall{ i \in \Lambda, d \in D_i }
\end{equation}
\vspace{-0.3cm}
\begin{equation}
\label{eq_27}
\small
\gamma_{id}^{DR} \leq \delta_{id}^{DR}, \quad\forall{ i \in \Lambda, d \in D_i }
\end{equation}
\vspace{-0.3cm}
\begin{equation}
\label{eq_28}
\small
\gamma_{id}^{DR} \geq \delta_{id}^{DR} - P_{id}^{DR,b}(1 - \omega_{id}^{DR}), \quad\forall{ i \in \Lambda, d \in D_i }
\end{equation}
\vspace{-0.3cm}
\begin{equation}
\label{eq_29}
\small
\gamma_{id}^{DR} \geq 0, \quad\forall{ i \in \Lambda, d \in D_i }
\end{equation}

The applied portion of DR block $d$ of each load at most can be equal to the capacity of step $d$. Furthermore, since the price of each step increases respectively, the DR blocks should be used accordingly. Mathematically discussing, if $\delta_{i(d+1)}^{DR}$ is non-zero, then $\delta_{id}^{DR}$ should be equal to $P_{id}^{DR,b}$. In other words, if the $d$th step of the DR contract of a load was used, all the previous steps would be filled due to their lower price. The mathematical formulation of these concepts is as follows:
\vspace{-0.03cm}
\begin{equation}
\label{eq_30}
\small
\delta_{id}^{DR} \geq \omega_{i(d+1)}^{DR} P_{id}^{DR,b}, \quad\forall{ i \in \Lambda, d \in D_i \backslash \{n\} }
\end{equation}

The loads with DR contracts are determined to be restored if they are fully restored, i.e., none of their offered DR blocks has been utilized.
\vspace{-0.03cm}
\begin{equation}
\label{eq_31}
\small
\omega_i^L \leq 1 - \omega_{id}^{DR}, \quad\forall{ i \in \Lambda, d \in D_i }
\end{equation}

\vspace{-0.3cm}
\begin{equation}
\label{eq_32}
\small
\omega_i^L \geq \sum_{d \in D_i}{ (1 - \omega_{id}^{DR}) - (n_d - 1) }, \quad\forall{ i \in \Lambda }
\end{equation}

Constraint (\ref{eq_31}) shows that if one of the DR blocks is used, the load is not restored, where constraint (\ref{eq_32}) represents that if none of the DR blocks is used, the load status is 1.

\vspace{-0.1cm}
\section{GBD-based Approach for CLR Problem}
\label{section_4}

\begin{figure}[t]
\centering
\includegraphics[width=0.98\linewidth]{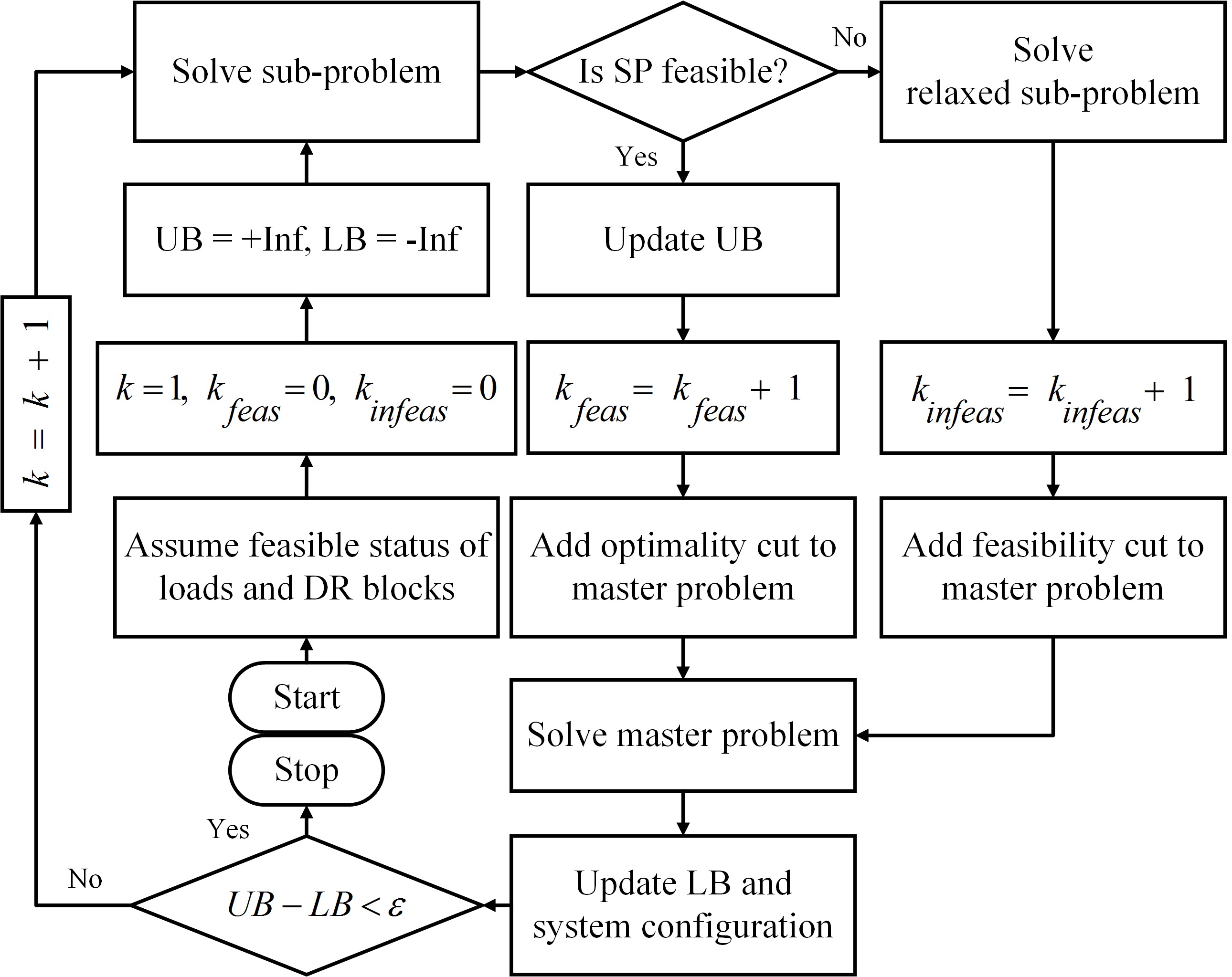}\vspace{-0.2cm}
\caption{Flowchart of two-step GBD framework to solve CLR}
\vspace{-0.5cm}
\label{fig_6}
\end{figure}

The CLR problem formulated in the second stage of the proposed method constitutes an MINLP optimization problem in which directly solving it needs highly computational efforts. Instead of considering all decision variables and constraints of the proposed MINLP simultaneously, the problem is solved based on GBD, which divides the problem into a master and sub-problem that are solved iteratively. The flowchart of the proposed two-step framework to solve CLR is illustrated in Fig. \ref{fig_6}. The integer variables, i.e., $\omega_i^L$ and $\omega_{id}^{DR}$, are determined in the master problem, which is an MILP problem, and the sub-problem, which is a convex NLP, solves the optimal power flow.

The master problem is formulated as follows:
\vspace{-0.03cm}
\begin{equation}
\label{eq_33}
\small
\min{\mu_B}
\end{equation}
Subject to constraints (\ref{eq_20}), (\ref{eq_31})-(\ref{eq_32}) and the optimality and feasibility cuts as follows:
\vspace{-0.03cm}
\begin{equation}
\label{eq_34}
\small
\begin{split}
\mu_B \geq f_{fs}^{(k-1)} + \sum_{i \in L}{ \lambda_{fs_i}^{(k-1)} (\omega_i^L - \omega_i^{L^{(k-1)}}) } + \quad\quad\quad \\
\sum_{i \in \Lambda}{ \sum_{d \in D_i}{ \mu_{fs_{id}}^{(k-1)} (\omega_{id}^{DR} - \omega_{id}^{{DR}^{(k-1)}}) } }, \forall{ k = 1,...,k_{fs} }
\end{split}
\end{equation}
\begin{equation}
\label{eq_35}
\small
\begin{split}
0 \geq f_{infs}^{(k-1)} + \sum_{i \in L}{ \lambda_{infs_i}^{(k-1)} (\omega_i^L - \omega_i^{L^{(k-1)}}) } + \quad\quad\quad\quad \\
\sum_{i \in \Lambda}{ \sum_{d \in D_i}{ \mu_{infs_{id}}^{(k-1)} (\omega_{id}^{DR} - \omega_{id}^{{DR}^{(k-1)}}) } }, \forall{ k = 1,...,k_{infs} }
\end{split}
\end{equation}
Where $\mu_B$ is the lower bound of the problem and $k_{fs}$ and $k_{infs}$ are the iteration counter of the feasible and infeasible sub-problem solutions, respectively. The value of the objective functions of the sub-problem and relaxed sub-problem in the previous iteration is shown by $f_{fs}^{(k-1)}$ and $f_{infs}^{(k-1)}$, respectively, and the Lagrangian multipliers associated with the coupling constraints of $\omega_i^L$ and $\omega_{id}^{DR}$ from the previous iteration of the sub-problem and relaxed sub-problem are denoted by $\lambda_{fs_i}^{(k-1)}$, $\mu_{fs_{id}}^{(k-1)}$, $\lambda_{infs_i}^{(k-1)}$, and $\mu_{infs_{id}}^{(k-1)}$, respectively. Constraints (\ref{eq_34})-(\ref{eq_35}) represent the optimality and feasibility cuts, respectively.

The objective function of the sub-problem is (\ref{eq_6}) subject to constraints (\ref{eq_7})-(\ref{eq_10}), (\ref{eq_12})-(\ref{eq_19}), (\ref{eq_21})-(\ref{eq_22}), (\ref{eq_25})-(\ref{eq_30}), and the following coupling constraints:
\vspace{-0.03cm}
\begin{equation}
\label{eq_36}
\small
\omega_i^L = \omega_i^{L^{(k)}} \quad : \lambda_{ij}^{(k)}, \quad\forall{ i \in L }
\end{equation}
\vspace{-0.3cm}
\begin{equation}
\label{eq_37}
\small
\omega_{id}^{DR} = \omega_{id}^{DR^{(k)}} \quad : \mu_{ij}^{(k)}, \quad\forall{ i \in \Lambda, d \in D_i }
\end{equation}

As it was stated earlier, it is noteworthy that in all above-mentioned formulation, $V_i^2$ and $I_{ij}^2$ will be replaced by $v_i$ and $i_{ij}$, respectively.

If the sub-problem is infeasible, the relaxed sub-problem will be solved. The objective function of the relaxed sub-problem is as follows:
\vspace{-0.03cm}
\begin{equation}
\label{eq_38}
\small
\min{\zeta}
\end{equation}
Subject to the coupling constraints (\ref{eq_36})-(\ref{eq_37}) and relaxed version of the sub-problem equations in which $\zeta$ is a non-negative slack variable that is used to relax the constraints of the sub-problem.

Benders iterative procedure is continued until the convergence criterion is met. At each iteration, the upper bound of the problem is decreased by the sub-problem, and the lower bound is increased by the master problem. Once the upper and lower bounds reach each other, the iterative procedure stops.

\vspace{-0.1cm}
\section{Case Studies}
\label{section_5}
In this study, two test cases, i.e., modified 33-bus system and modified 69-bus system are employed to evaluate the effectiveness of the proposed critical load restoration method by coordinating microgrids and demand response programs.

\begin{figure}[t]
\centering
\includegraphics[width=0.98\linewidth]{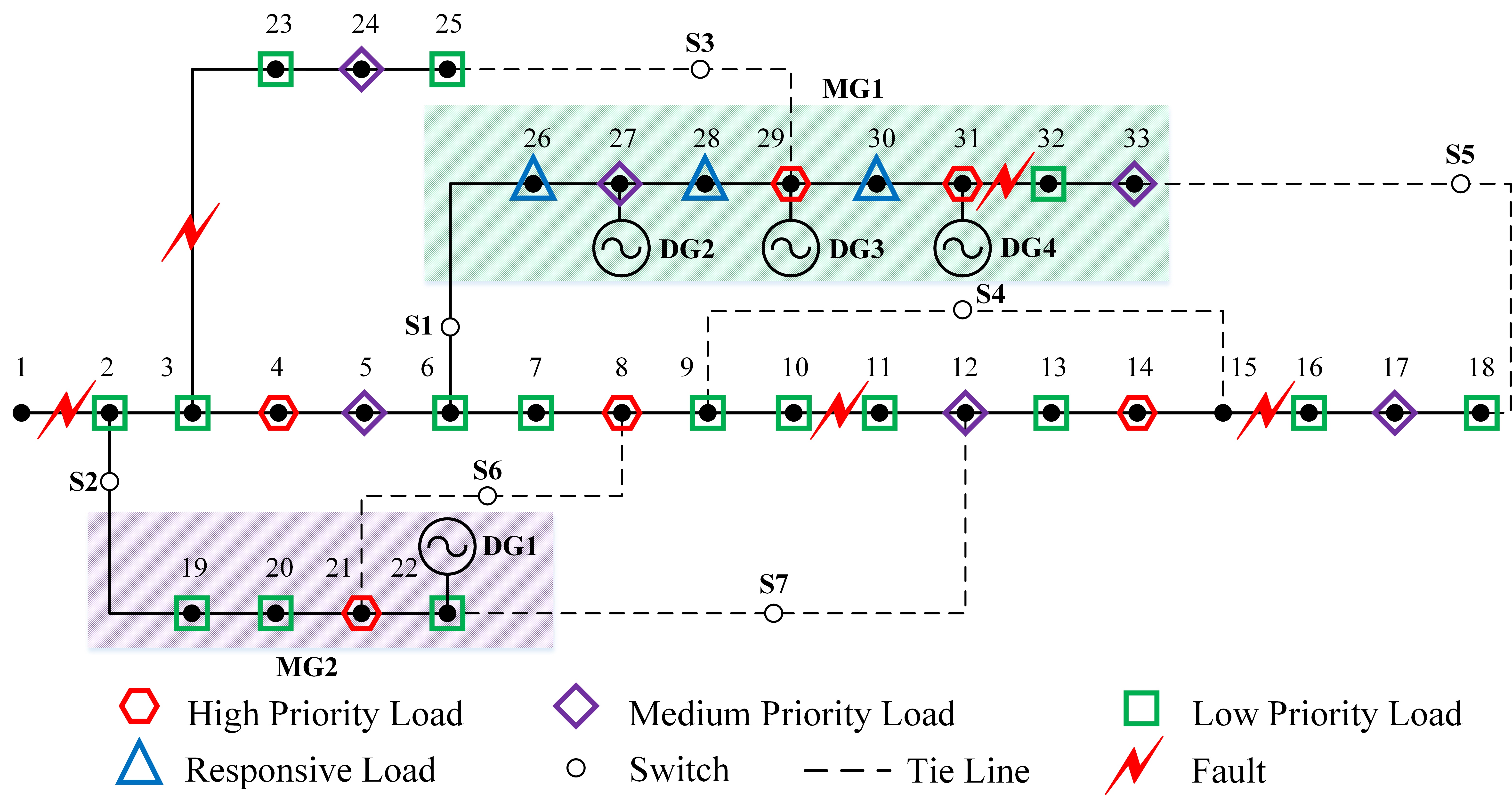}\vspace{-0.2cm}
\caption{Modified 33-bus system}
\vspace{-0.5cm}
\label{fig_7}
\end{figure}

\begin{table}[b]
\vspace{-0.5cm}
\centering
\caption{DGs Data in 33-Bus System}
\label{tab_1}
\vspace{-0.3cm}
\resizebox{0.95\linewidth}{!}{%
\begin{tabular}{ccccc}
\toprule
Generator & Microgrid & Bus  & $Q_i^{DG,Max}$(kVAr) & $S_i^{DG,Max}$(kVA) \\ \midrule
DG1 & MG2 & 22 &  50 & 100 \\ 
DG2 & MG1 & 27 & 450 & 630 \\ 
DG3 & MG1 & 29 & 300 & 425 \\ 
DG4 & MG1 & 31 & 220 & 300 \\ \bottomrule
\end{tabular}%
}
\end{table}

\begin{table}[b]
\vspace{-0.3cm}
\centering
\caption{Demand Response Results in 33-Bus System}
\label{tab_2}
\vspace{-0.3cm}
\resizebox{0.75\linewidth}{!}{%
\begin{tabular}{cccc}
\toprule
Bus & Microgrid & Load (kW)  & Status of DR Blocks \\ \midrule
26 & MG1 &  60 & (1, 1, 1, 0) \\ 
28 & MG1 &  60 & (1, 1, 0, 0) \\ 
30 & MG1 & 200 & (1, 1, 1, 1) \\ \bottomrule
\end{tabular}%
}
\end{table}

\vspace{-0.2cm}
\subsection{Case I: Modified 33-Bus System}
\label{section_5_1}
The first test case is the modified 33-bus system \cite{25627}. The modified 33-bus system is shown in Fig. \ref{fig_7}. There are two MGs in the distribution system. MG1 is a private MG whose loads have DR contracts and participates in critical load restoration. Also, MG2 participates in CLR, but since it is not private, there are no DR contracts. The microgrids are connected to the rest of the distribution system with switches $s_1$ and $s_2$. There are five tie lines in the distribution system. MG1 includes three DGs, and MG2 contains one DG. The DGs' information and data are represented in Table \ref{tab_1}. Assume that there are three types of loads in the distribution system, i.e., critical or high priority, medium, and low priority loads. The priority factor for high, medium, and low priority loads are 100, 10, and 0.1, respectively. The low priority loads in MG1 at nodes 26, 28, and 30 are supposed to offer four DR blocks, each of which is 25 percent of the total load. In other words, the DSO can utilize all the DR blocks of the load and down the total load if needed. The priority factors of the DR blocks are 0.3, 0.4, 0.5, 0.6, respectively.

Assume that after the occurrence of a disaster, five lines are faulted, as indicated in Fig. \ref{fig_7}. By using the proposed method, there is one target island as an electrical island. Available five tie lines make it possible to connect the separate areas and form an electrical island, as shown in Fig. \ref{fig_7}.

The proposed method in the first stage is utilized to determine the post-disaster topology. Considering weighted loads and DGs in the electrical island, the tie lines between the nodes 9-15 and 25-29 should be energized, as stated in the proposed method. Since both MGs are participating in the critical load restoration, their switches, i.e., $s_1$ and $s_2$, should be closed.

Three possible cases are studied, and the results are compared. The cases are i) coordinating MGs using DR contracts, ii) coordinating MGs without DR contracts, and iii) without coordinating MGs.

\begin{figure}[t]
\centering
\includegraphics[width=0.98\linewidth]{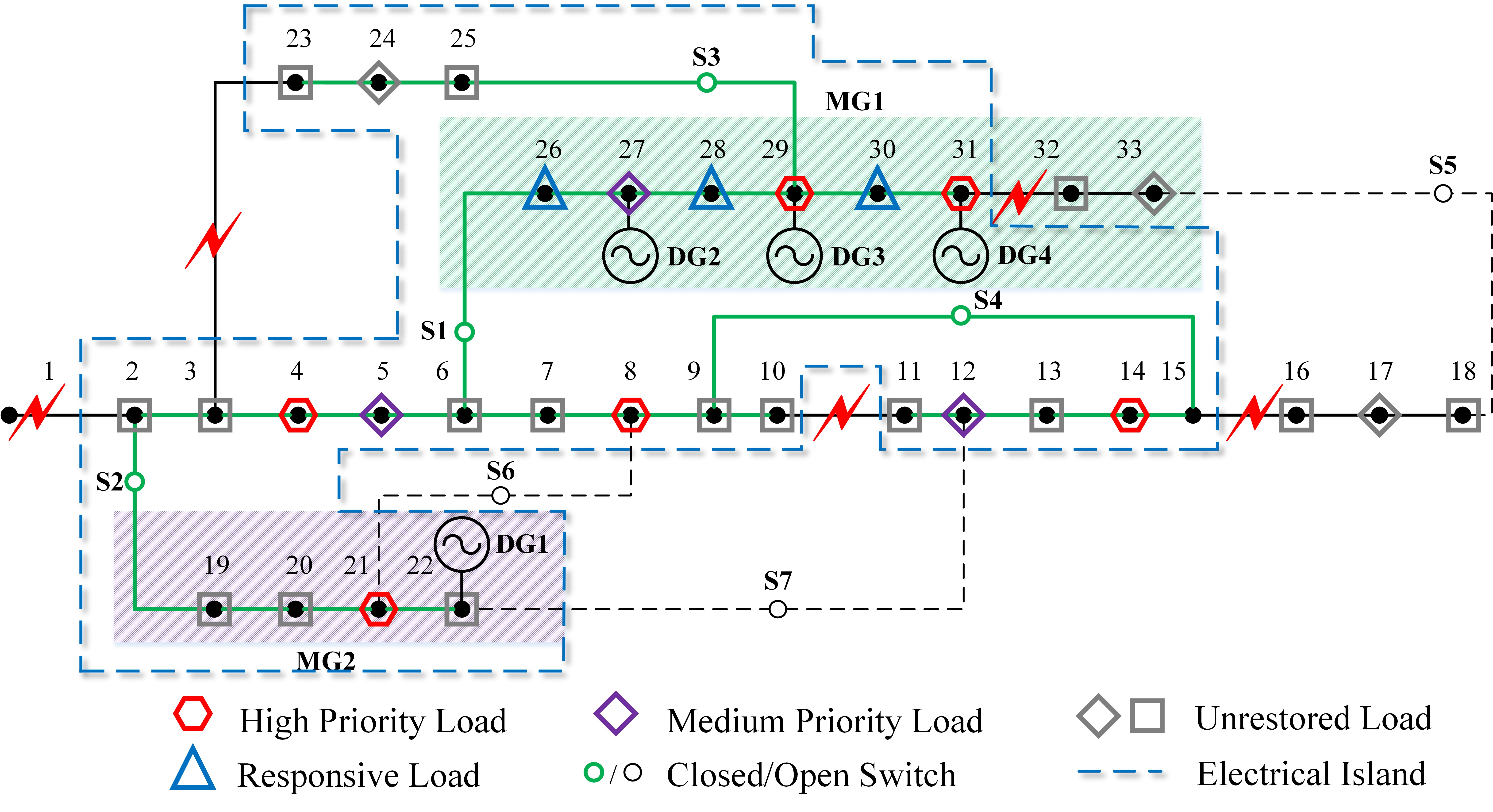}\vspace{-0.2cm}
\caption{Coordinating MGs in CLR with DR programs in 33-bus system}
\vspace{-0.2cm}
\label{fig_8}
\end{figure}

\begin{figure}[t]
\centering
\includegraphics[width=0.98\linewidth]{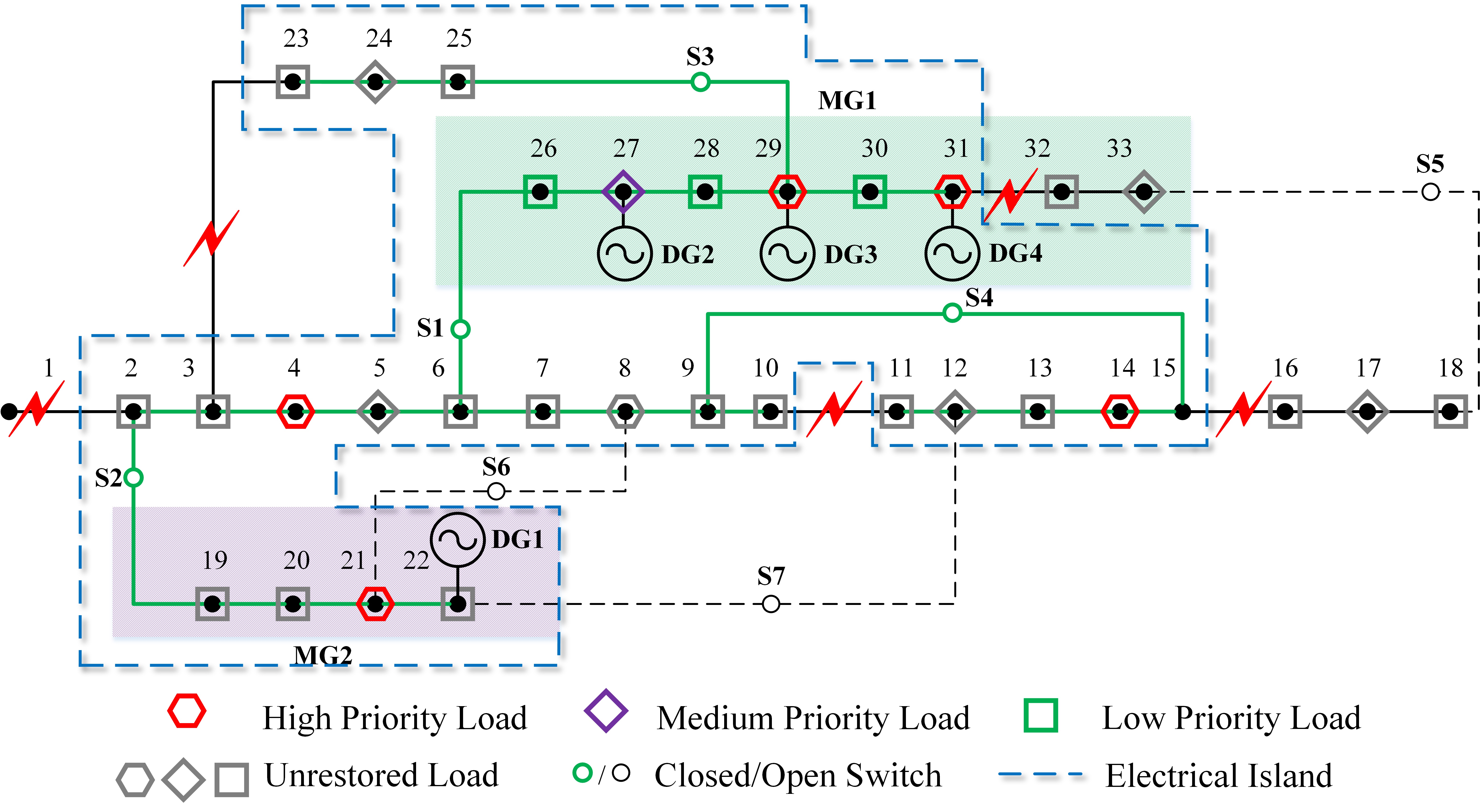}\vspace{-0.2cm}
\caption{Coordinating MGs in CLR without DR programs in 33-bus system}
\vspace{-0.5cm}
\label{fig_9}
\end{figure}

After applying the proposed method in the second stage, the restored loads are illustrated in Fig. \ref{fig_8}. The loads located in nodes 27, 29, and 31 belong to MG1, which is a private microgrid, are restored since they do not offer any DR contracts. By coordinating the resources of both microgrids, the high priority loads at buses 4, 8, 14, and 21 and medium priority loads at buses 5 and 12 are restored. Demand response results are shown in Table \ref{tab_2}.

To compare the results with the case in which there are no DR contracts, assume that the loads at buses 26, 28, and 30 do not offer any DR contracts. Thus, they should be restored like the other loads of the private MG. In this case, the critical loads at buses 4, 14, and 21 and none of the medium priority loads are restored, as illustrated in Fig. \ref{fig_9}. Therefore, as the results indicate, without DR contracts, the critical load at bus 8 and the medium priority loads at buses 5 and 12 will not be restored.

The third case is considering the post-event condition without microgrids’ coordination. If there were no coordination between the microgrids, the switches of both microgrids would be open, resulting in losing all the loads in the rest of the distribution system. Consequently, since there is no DG outside the MGs, all the loads which do not belong to MGs will be lost. The only restored loads are loads of MG1.

\begin{figure}[t]
\centering
\includegraphics[width=0.98\linewidth]{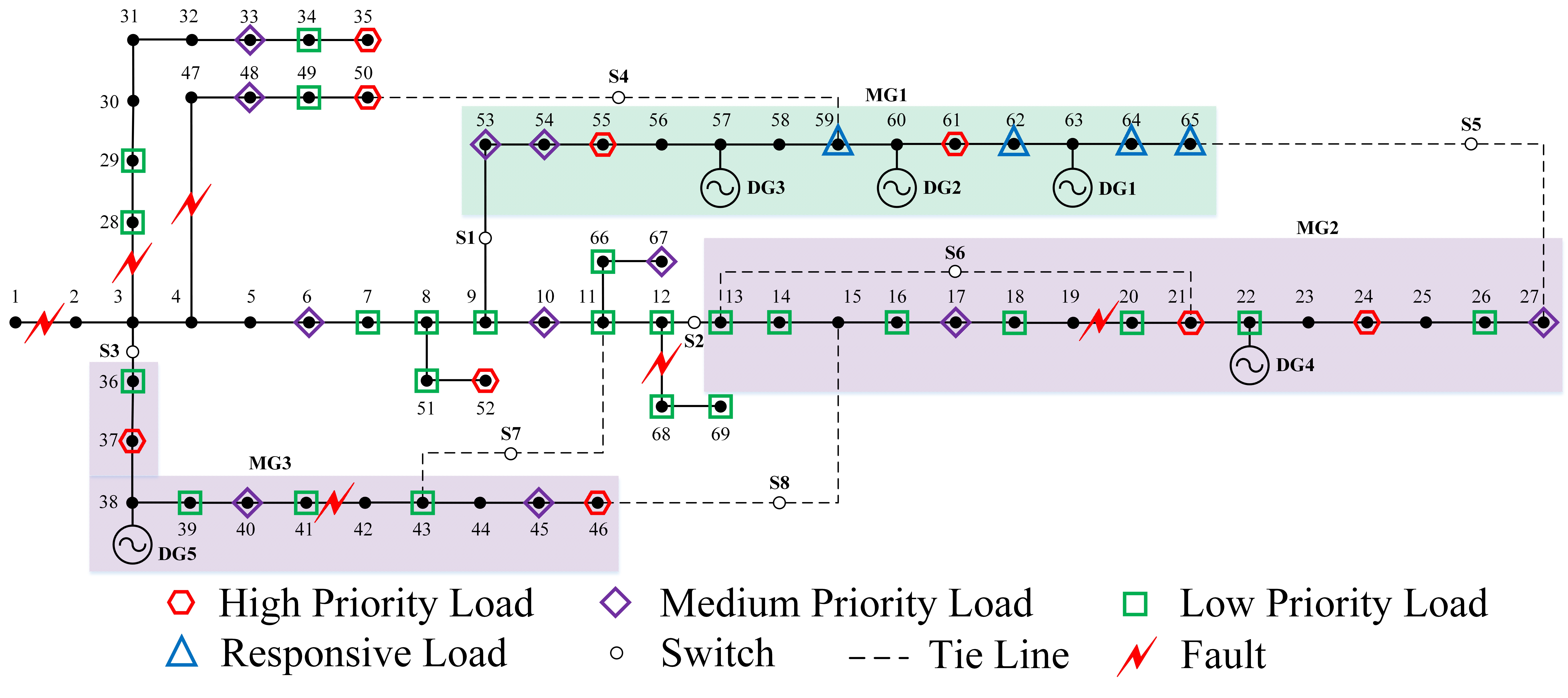}\vspace{-0.2cm}
\caption{Modified 69-bus system}
\vspace{-0.5cm}
\label{fig_10}
\end{figure}

The minimum voltage and power losses of the pre-disaster condition and the case in which the MGs are coordinated in the proposed CLR with DR programs can be compared. The minimum voltage of the network in the pre-disaster condition and after the occurrence of the incident is 0.9606 and 0.9840 per unit, respectively. Also, the power losses of the network in the normal and post-disaster operations are 140.65 kW and 7.23 kW, respectively. In the post-disaster condition, a few loads are supplied, resulting in higher minimum voltage and lower power losses.

As the results show, in this test case, coordinating MGs with using demand response programs causes to restore 100\% of critical loads and 50\% of medium priority loads. However, coordinating MGs without utilizing DR programs yields to restore 83\% of critical loads and 16\% of medium priority loads. In the third case, in which there is no coordination between microgrids, 33\% of high priority loads and 16\% of medium priority loads are restored. Consequently, in this case study, coordinating MGs with and without applying DR programs increases the restoration of the critical loads by 67\% and 50\%, respectively.

\begin{table}[b]
\vspace{-0.5cm}
\centering
\caption{DGs Data in 69-Bus System}
\label{tab_3}
\vspace{-0.3cm}
\resizebox{0.95\linewidth}{!}{%
\begin{tabular}{ccccc}
\toprule
Generator & Microgrid & Bus  & $Q_i^{DG,Max}$(kVAr) & $S_i^{DG,Max}$(kVA) \\ \midrule
DG1 & MG1 & 63 & 800 & 1130 \\ 
DG2 & MG1 & 60 & 500 & 700 \\ 
DG3 & MG1 & 57 & 500 & 700 \\ 
DG4 & MG2 & 22 & 100 & 150 \\
DG5 & MG3 & 38 & 50 &  70 \\ \bottomrule
\end{tabular}%
}
\end{table}

\begin{table}[b]
\vspace{-0.3cm}
\centering
\caption{Demand Response Results in 69-Bus System}
\label{tab_4}
\vspace{-0.3cm}
\resizebox{0.75\linewidth}{!}{%
\begin{tabular}{cccc}
\toprule
Bus & Microgrid & Load (kW)  & Status of DR Blocks \\ \midrule
59 & MG1 & 100 & (1, 1, 1, 1) \\ 
62 & MG1 &  32 & (1, 1, 1, 1) \\ 
64 & MG1 & 227 & (1, 1, 1, 1) \\
65 & MG1 &  59 & (1, 1, 1, 1) \\ \bottomrule
\end{tabular}%
}
\end{table}

\vspace{-0.2cm}
\subsection{Case II: Modified 69-Bus System}
\label{section_5_2}
The second test case is the modified 69-bus system \cite{4302553}, which is illustrated in Fig. \ref{fig_10}. The modified 69-bus system includes three MGs in which MG1 is a private MG with DR contracts with its customers. MG2 and MG3 are not private MGs. Therefore, there are no DR contracts with their customers. MG1 contains three DGs and each of MG2 and MG3 includes one DG. The DGs' data is shown in Table \ref{tab_3}. The distribution system possesses five tie lines, which can be used in the reconfiguration process. The assumptions for the loads, DR blocks, and their priority factors are the same as the first test case.

\begin{figure}[t]
\centering
\includegraphics[width=0.98\linewidth]{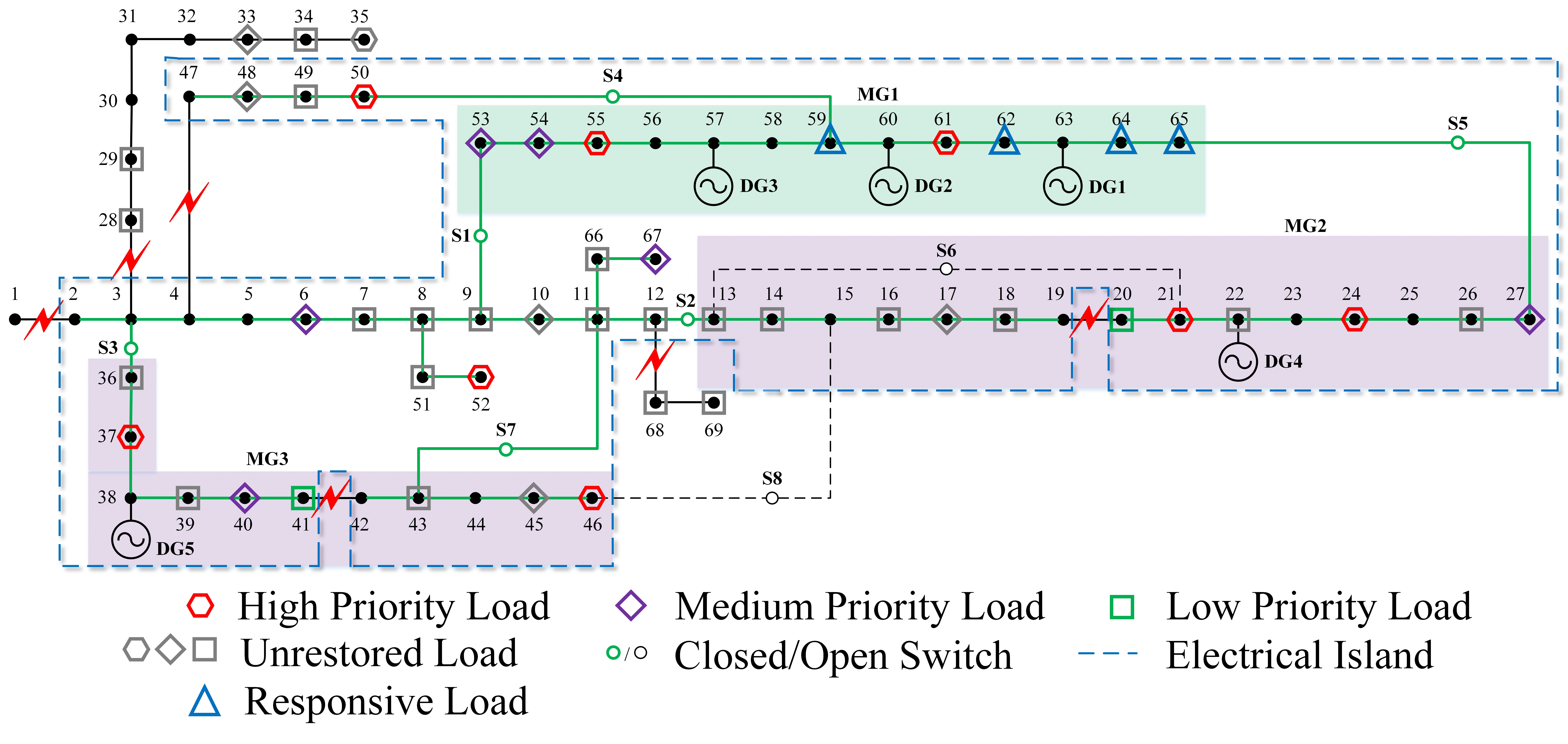}\vspace{-0.2cm}
\caption{Coordinating MGs in CLR with DR programs in 69-bus system}
\vspace{-0.5cm}
\label{fig_11}
\end{figure}

After a major event, six lines are faulted as represented in Fig. \ref{fig_10}. First, one target island is detected considering the available lines and tie lines. At the first stage of the proposed method, the post-event topology of the network is determined as indicated in Fig. \ref{fig_11}. All MGs are willing to participate in the CLR program, so, their switches would be closed. As shown in Fig. \ref{fig_11}., $s_1$, $s_2$, $s_3$, $s_4$, $s_5$, and $s_7$ are closed to make an electrical island and the post-disaster topology.

\begin{figure}[t]
\vspace{+0.3cm}
\centering
\includegraphics[width=0.98\linewidth]{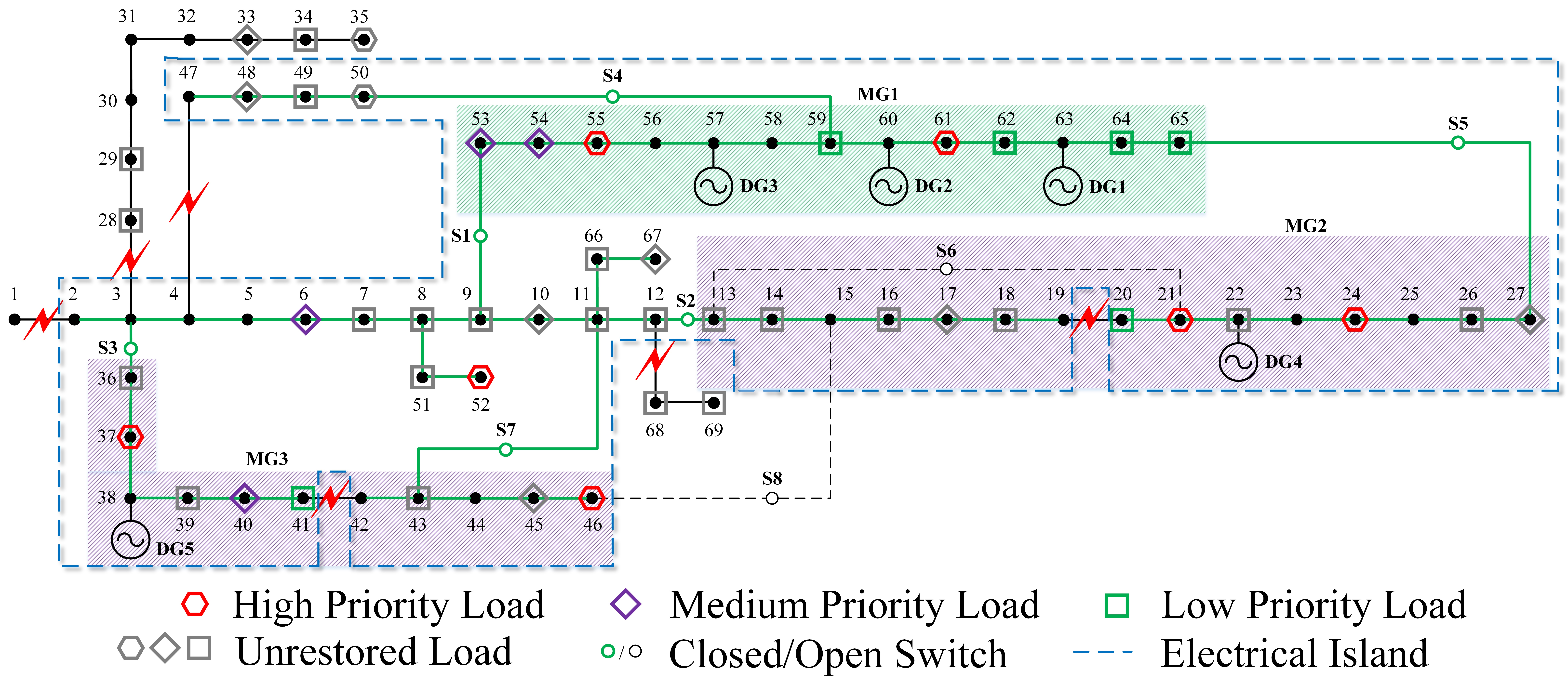}\vspace{-0.2cm}
\caption{Coordinating MGs in CLR without DR programs in 69-bus system}
\vspace{-0.5cm}
\label{fig_12}
\end{figure}

Three possible cases, i.e., i) coordinating MGs using DR contracts, ii) coordinating MGs without DR contracts, and iii) without coordinating MGs are evaluated to compare the results. In the first case, i.e., coordinating MGs using DR contracts, eight critical loads and six medium priority loads are restored as illustrated in Fig. \ref{fig_11}. The status of the DR blocks for the responsive loads are represented in Table \ref{tab_4}. In the second case, i.e., coordinating MGs without DR contracts, seven critical loads and four medium priority loads are restored as shown in Fig. \ref{fig_12}. In the third case, i.e., without coordinating MGs, the switches of the MGs are opened and each MG supplies its own critical loads with its available DGs resulting in four critical loads and three medium priority loads being restored.

According to the results, in the first case, i.e., coordinating MGs using DR contracts, 89\% of the critical loads and 55\% of the medium priority loads are restored while in the second case, i.e., coordinating MGs without DR contracts, 78\% of the critical loads and 36\% of the medium priority loads are restored. Also, in the third case in which there is no coordination between MGs, 44\% of the critical loads and 27\% of the medium priority loads are restored. Therefore, the MGs coordination with and without using DR contracts increases the number of restored critical loads by 45\% and 34\%, respectively.

\vspace{-0.1cm}
\section{Conclusion and Future Work}
\label{section_6}
In this paper, a two-stage decision-making critical load restoration strategy was proposed. The main contributions of the paper are proposing a new method to determine the post-disaster topology, utilizing demand response as a resource in the presence of emergency conditions to restore the critical loads of the distribution system, and considering the ownerships of the microgrids in the distribution system.

In the proposed method, at the first stage, after determining the possible target islands as electrical islands, the topology of each electrical island is determined. Then, the second stage determines the amount of the restored loads, the outputs of the distributed generators, and the utilized demand response programs. Additionally, a two-step GBD-based algorithm is proposed to solve the second stage of the proposed method.

Since renewables such as photovoltaic generators are intermittent sources, there are uncertainties in their generations, resulting in uncertainties in microgrids’ capacities. Although, as mentioned in the benefits of the coordination of microgrids, managing uncertainties of the renewables are enhanced by the coordinated performance of the microgrids, an accurate model of uncertainties of renewables is required to thoroughly investigate the impact of uncertainties in the coordination of microgrids in the critical load restoration strategy \cite{8283770}. Additionally, since the proposed algorithm is decision-making, modeling and solution techniques for optimization problems with random variables are needed \cite{7447808}. Both issues are challenging, and in order to comprehensively deal with uncertainties in the proposed CLR using MGs coordination and responsive loads, further research is demanded.

\ifCLASSOPTIONcaptionsoff
  \newpage
\fi
	\bibliographystyle{IEEEtran}
	\bibliography{main}
\begin{IEEEbiography}[{\includegraphics[width=1in,height=1.25in,clip,keepaspectratio]{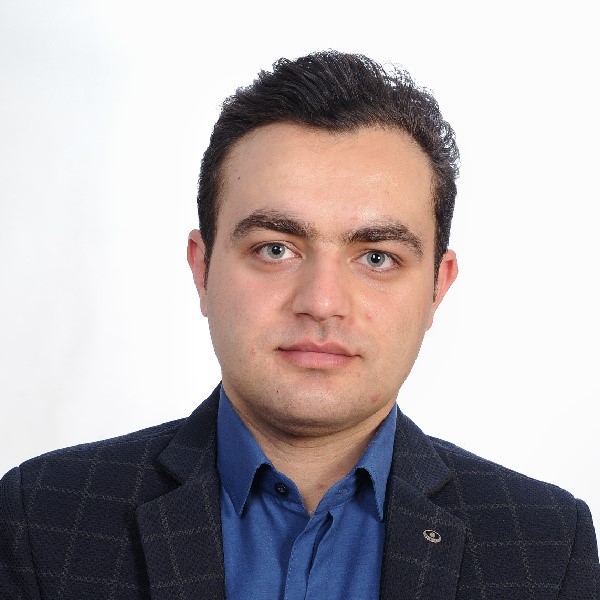}}]{Ali Shakeri Kahnamouei} (S'19) received the B.Sc. degree in Electrical Engineering from University of Tabriz, Tabriz, Iran, in 2014, and the M.Sc. degree from Tarbiat Modares University, Tehran, Iran, in 2017. He is currently working toward the Ph.D. degree at the School of Electrical Engineering and Computer Science, Washington State University, Pullman, WA, USA. His current research interests include power system protection and control and power system resilience.
 \end{IEEEbiography}
  \vspace{-1cm} \begin{IEEEbiography}[{\includegraphics[width=1in,height=1.25in,clip,keepaspectratio]{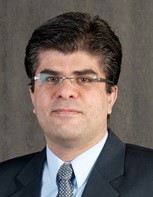}}]{\textbf{Saeed Lotfifard}} (S'08-M'11-SM'17) received his Ph.D. degree in electrical engineering from Texas A\&M University, College Station, TX, in 2011. Currently, he is an associate professor at Washington State University, Pullman. His research interests include protection, control and operational security of inverter-based power grids. Dr. Lotfifard is an Editor for the IEEE Transactions on Smart Grid and IEEE Transactions on Sustainable Energy.
\end{IEEEbiography}
\end{document}